\documentclass[preprint2]{emulateapj} 
\usepackage{amsmath}
\usepackage{natbib}
\usepackage{graphicx}
\usepackage{epsf}
\usepackage{color}
\usepackage{ulem}
\input{epsf}
\usepackage{epsfig}
\DeclareGraphicsExtensions{.jpg,.pdf,.png,.eps,.ps}
\graphicspath{{FIGURES/}}

\begin{document}

\title{Cross-correlation of Far-Infrared background anisotropies and CMB Lensing\\ from Herschel and Planck satellites}

\author{Ye Cao$^{1,2}$, Yan Gong$^{1*}$, Chang Feng$^{3,4}$, Asantha Cooray$^{4}$, Gong Cheng$^{5,2}$, Xuelei Chen$^{5,2,6}$}

\affil{$^{1}$ Key Laboratory of Space Astronomy and Technology, National Astronomical Observatories,\\ Chinese Academy of Sciences, Beijing 100012, China}
\affil{$^{2}$ School of Astronomy and Space Sciences, University of Chinese Academy of Sciences, Beijing 100049, China}
\affil{$^{3}$ Department of Physics, University of Illinois at Urbana-Champaign, 1110 W Green St, Urbana, IL, 61801, USA}
\affil{$^{4}$ Department of Physics and Astronomy, University of California, Irvine, CA 92697, USA}
\affil{$^{5}$ Key Laboratory of Computational Astrophysics, National Astronomical Observatories,\\ Chinese Academy of Sciences, Beijing 100012, China}
\affil{$^{6}$ Center for High Energy Physics, Peking University, Beijing 100871, China}

 
\begin{abstract}

The cosmic infrared background (CIB) anisotropies and cosmic microwave background (CMB) lensing are powerful measurements for exploring the cosmological and astrophysical problems. In this work, we measure the auto-correlation power spectrum of the CIB anisotropies in the Herschel-SPIRE HerMES Large Mode Survey (HeLMS) field, and the cross power spectrum with the CMB lensing measurements from the Planck satellite. The HeLMS field covers more than 270  deg$^2$, which is much larger than previous analysis.  We use the Herschel Level 1 time stream data to merge the CIB maps at 250, 350, and 500 $\rm{\mu m}$ bands, and mask the areas where the flux is greater than ${3\sigma\ (\sim50\rm\ mJy/beam)}$ or no measured data. We obtain the final CIB power spectra at $100\le\ell \le20000$ by considering several effects, such as beam function, mode coupling, transfer function, and so on. We also calculate the theoretical CIB auto- and cross-power spectra of CIB and CMB lensing by assuming that the CIB emissivity follows Gaussian distribution in redshift. We find that, for the CIB auto power spectra, we obtain the signal to noise ratio (SNR) of 15.9, 15.7, and 15.3 at 250, 350, and 500 $\mu$m, and for the CIBxCMB lensing power spectra, SNR of 7.5, 7.0, and 6.2 at 250, 350, and 500 $\mu$m, respectively. Comparing to previous works, the constraints on the relevant CIB parameters are improved by factors of 2$\sim$5 in this study.
\end{abstract}


\bigskip\bigskip

\section{Introduction}
The cosmic infrared background (CIB) anisotropies survey is a basic observation for modern astronomical research. Through the studies of CIB, we can obtain important information to help resolving cosmological and astrophysical problems, such as the formation and evolution of the large-scale structure of the Universe, the star formation history, epoch of reionization, and so on. The first direct CIB measurements were performed by the FIRAS and DIRBE instrument of the COBE satellite \citep{Puget96,Fixsen98,Hauser98}. In recent years, a number of increasingly sensitive experimental measurements of CIB have made milestone contributions, such as Herschel and Planck space telescope, which dedicate to measuring the cosmic far-infrared background \citep{Griffin10, Planck11I}. These observations with increased resolution, sensitivity, frequency coverage and detection area provide a powerful constraints on the CIB models \citep{Planck14XXX, Serra14, Mak17, Lenz19}. According to current models, the most important component of the CIB is the infrared emission from unresolved dusty star-forming galaxies \citep{Puget96,Lagache03,Dole04}, which have a redshift distribution peaked from redshift $z \sim1$ to 2 \citep[e.g.][]{Bethermin12,Schmidt15}.

On the other hand, the effect of gravitational lensing was first considered by \cite{Blanchard87}. In the proceeding years, a method for accurately estimating the effects of lensing has been proposed in the $\Lambda \mathrm{CDM}$ framework \citep[e.g.][]{Challinor05}. In recent years, a number of experiments for measuring CMB lensing with high resolution and high sensitive have been performed, e.g. WMAP and Planck missions \citep{Hirata04,Planck14XVII}. The direction of photons coming from the last scattering surface can be changed by the gravitational potential \citep{Okamoto03,Lewis06}. Shear and magnification are the most important effects from gravitational lensing in the observed fluctuations. The cosmological parameters, the properties of dark matter and dark energy, and other astrophysical effects can be derived by analyzing the lensing signal. 

In this work, we measure the auto- and cross-power spectra of the CIB from Herschel and CMB lensing from Planck at 250, 350, and 500 $\rm \mu$m. We analyze the Herschel-SPIRE HerMES Large Mode Survey (HeLMS) field, which covers more than 270 deg$^2$ and is much larger than previous similar analysis by about one order of magnitude \citep[e.g.][]{Thacker13,Thacker15,Viero13}. This can significantly pin down the derived statistical errors of the CIB power power spectrum, and let us explore larger scales to extract more CIB information. We use Madmap algorithm to merge the the Herschel Level 1 time stream data as the CIB maps using the Herschel Interactive Processing Environment (HIPE) \citep{Ott10}, and remove the contamination from bright sources with a $3\sigma$ flux cut. The CMB lensing data is obtained form the Planck 2013 data release \citep{Planck11I}. The estimation and correction of the auto- and cross-power spectrum is based on the work of \cite{Cooray12}. We then obtain the auto-power spectrum of CIB anisotropies from $100 \leq \ell \leq 20,000$ and the cross-power spectrum of CIB and CMB lensing from $100\leq \ell \leq 2000$. Besides, we also calculate the theoretical model and perform a Markov chain Monte Carlo (MCMC) analysis to constrain the free parameters, and derive the CIB mean emissivity as a function of redshift.

The paper is organized as follows: in Section \ref{sub:data}, we analyze the Herschel data obtained by HIPE and Planck data from Planck Legacy Archive. The mask generation process is also discussed. In Section \ref{sub:spectrum}, we describe the power spectrum measurements and corrections. The final auto- and cross-power spectra, and the errors are shown in \S\ref{subsec:final_power}. In Section \ref{model}, we describe the theoretical model and fit the data by MontePython code \citep{Audren13}. We finally summarize the results in Section \ref{conclusions}.
In this work, we adopt the standard $\Lambda \mathrm{CDM}$ cosmological model with parameter values derived from \cite{Planck18VI}, which gives $h$ = 0.674, $\Omega_{\mathrm{m}}$ = 0.315, $\Omega_{\mathrm{b}} h^2$ = 0.0224, $\sigma_{8}$ = 0.811, and $n_s$ = 0.965.

\section{Data analysis}
\label{sub:data}

\subsection{CIB maps from Herschel}
\label{subsec:Herschel}
In the Herschel data, we make use of the public Herschel-SPIRE HerMES Large Mode Survey (HeLMS, {PI: Marco Viero}) data that is obtained from Herschel Science Archive\footnote{\tt http://archives.esac.esa.int/hsa/whsa/}. As the largest area observed in HerMES field, HeLMS spans $348^\circ < {\rm RA} < 20^\circ$ and $-9^\circ < {\rm Dec} < +9^\circ$, and covers more than 270 deg$^2$ of the sky with the mean flux density of the Galactic cirrus is 1.2 $\rm{MJy/Sr}$. The HeLMS field involves 11 independent tiles, and each tile contains 41 scan-lines. In addition to the large area, the advantage of the HerMES field is that almost each area is observed twice in two nearly orthogonal scanning directions, which can help us to analyze the instrumental noise power spectrum. 

 \begin{figure*}
\centering
\includegraphics[width=2.1\columnwidth]{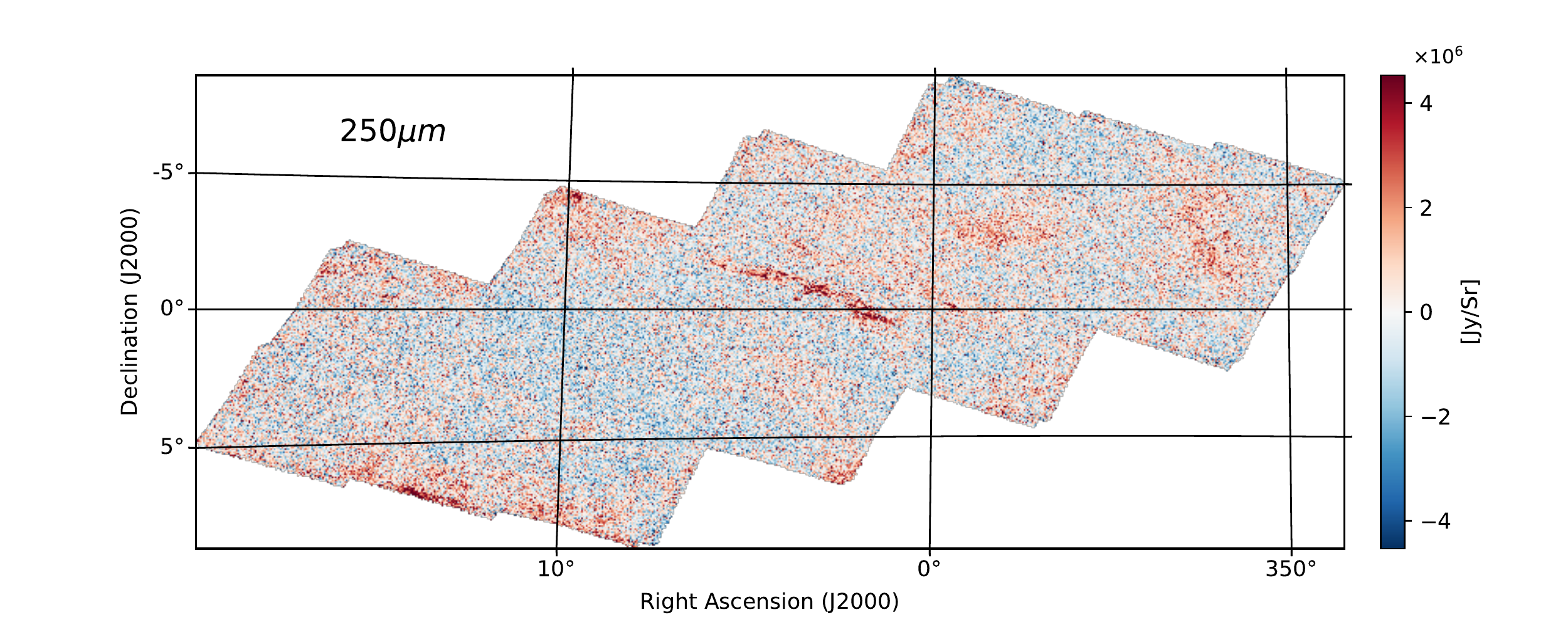}
\includegraphics[width=2.1\columnwidth]{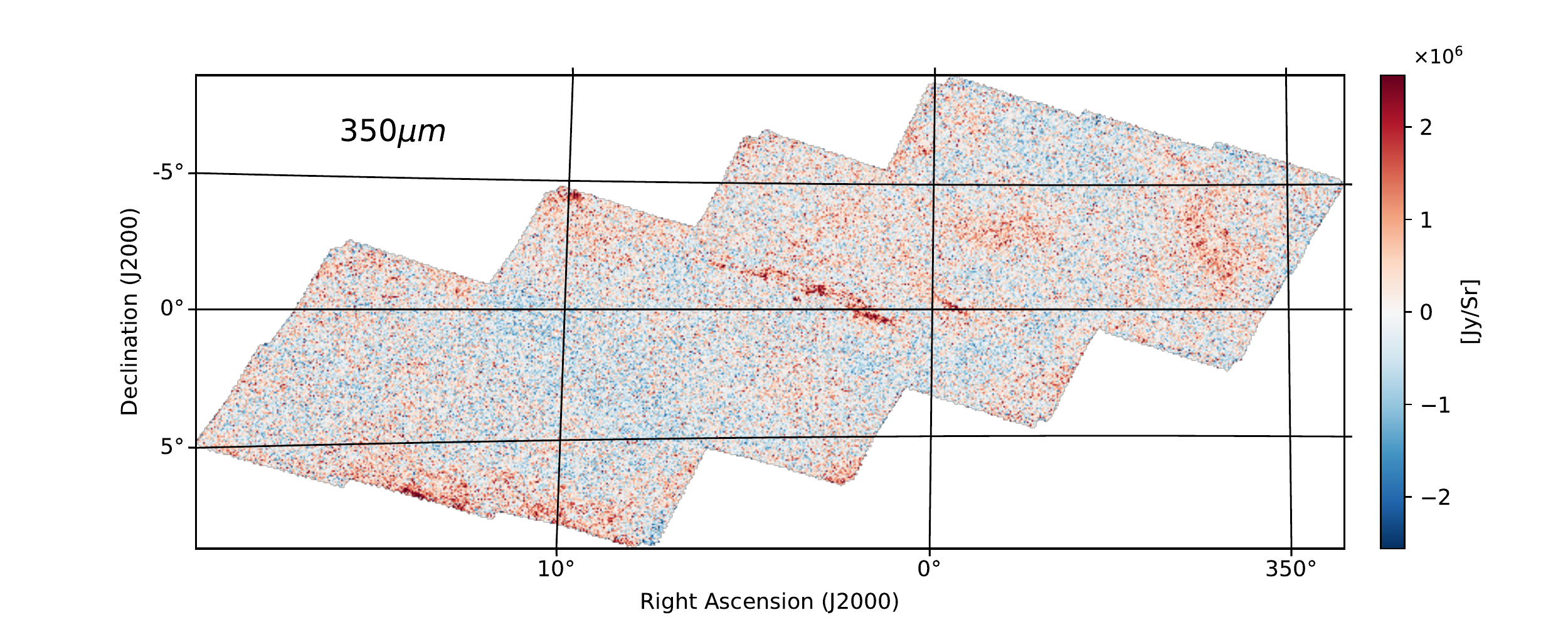}
\includegraphics[width=2.1\columnwidth]{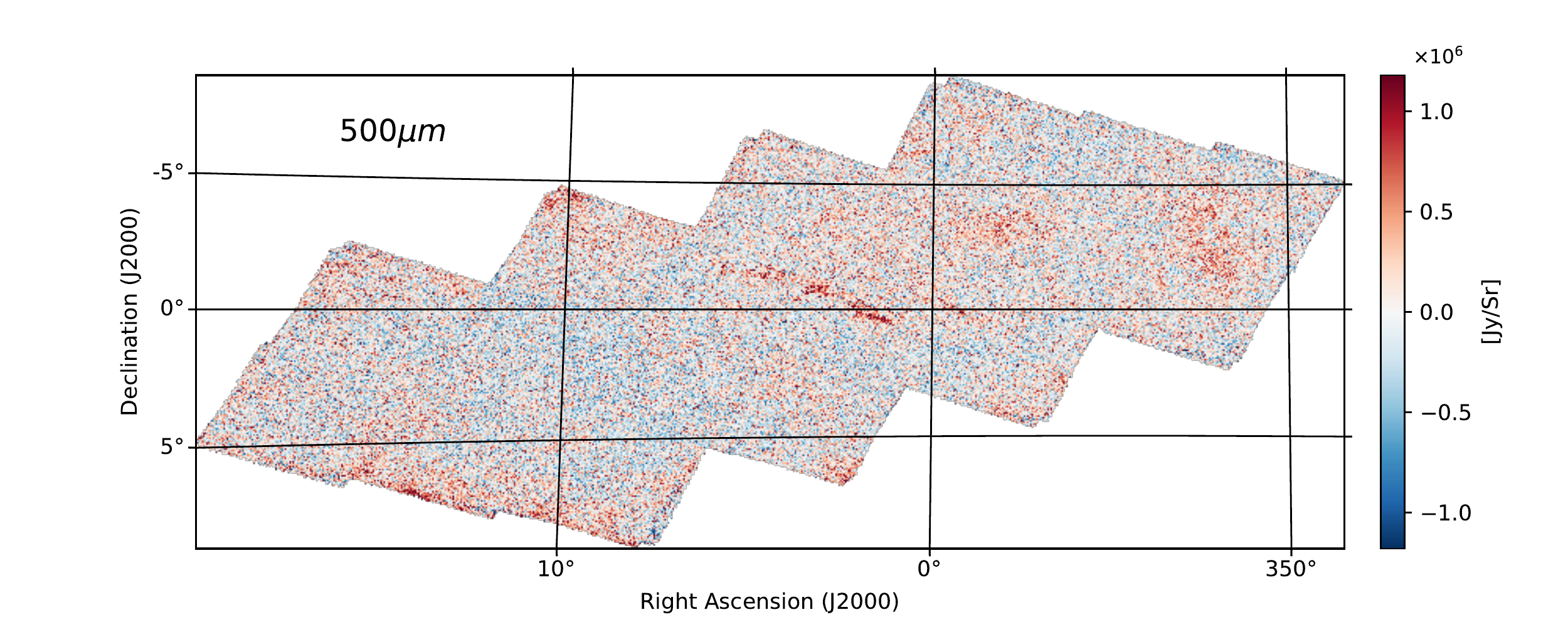}
\caption{The final Herschel-SPIRE HeLMS maps at 250 (top panel), 350 (middle panel), and 500$\rm{\mu m}$ (bottom panel) merged by HIPE. The colored areas are the scanned sky areas, which are scanned twice in two nearly orthogonal directions. Every map is merged with 11 independent tiles (5 laterally scanned tiles and 6 longitudinally scanned tiles). These CIB maps are calibrated by extended sources, and their units have been converted from $\rm{Jy/beam}$ to $\rm{Jy/Sr}$. }
\label{fig:map_cib}
\end{figure*}

In this work, the CIB maps are generated using the Level 1 time stream data from all 11 Astronomical Observation Requests, and the observation IDs are 1342234749, 1342236232, 1342236234, 1342236240, 1342237550, 1342237553, 1342237563, 1342238251, 1342246580, 1342246632 and 1342247216. There are many stripes on the raw map produced by simply projecting the Level 1 time stream data onto the nearest sky map pixel. In order to eliminate this stripes, we need to perform a three-step process in each of the SPIRE bands before generating the final map.
First, Herschel-SPIRE detectors are only sensitive to relative variations \citep{Bernard10}, which results in the unknown absolute brightness of the measured filed. Therefore, the calibration values between the observed values and the actual values of the time stream data are different. This effect will be passed to the final map through the map making pipeline. We assume that the difference in the median of adjacent areas is negligible. Then we subtract the median value for each time stream data, and construct a tile with zero median. This works well for the overall flat background, and we will show the details of the effect of removing differences in median of all time streams in \S\ref{subsec:trans_func}.
Next, we apply the Polynomial Baseline Subtraction, reducing the effect due to cooler temperature variations for each individual scan line. Observations taken after the cooler recycle can be affected by a temperature drift, and it is especially serious in a single scan of long duration. The large scale striping in the raw map is clearly obvious, and it is a constant drift along the scan direction. Therefore, we usually treat the deviation caused by cooler temperature drift as linear change with time. We use a linear polynomial as the offset function, and fit the function using the data on a single scan, and then use the Polynomial Baseline Subtraction algorithm to correct the time streams scan by scan. 
Besides, many stripes come from the telescope system itself. In Herschel telescope system, the thermal and electronic stability are limited, we can find that the residual offsets in the flux calibration are different  from one detector to another. We use the de-striper algorithm to iteratively update offsets until an optimal solution is found.

In order to create a stripe-free map for the structured regions. The merged final maps are obtained by using the Madmap algorithm, which is a maximum-likelihood-estimate method for generating a final map \citep{Cantalupo10}. The Madmap method uses an iterative approach to minimize the Fourier transform of the map, and adopts the spectrum of the photometer channel noise to correct the final map. In this work, we set the pixel scale of merged final maps to be $6''$, $10''$ and $14''$ for the 250, 350, and 500 $\rm{\mu m}$ bands, respectively, and they are about 1/3 full-width-half-max (FWHM) of the beams. During the merger process, we also correct the effects of bolometer signal jumps , thermistor jumps and so on. 

To perform the flux calibration we make use of the method used in \cite{Griffin13}, and create the calibrated maps using the calibration tree with version 14.3. In addition to the statistical errors of measurement, the SPIRE flux calibration still has three sources of uncertainties, i.e. a systematic uncertainty in the flux calibration (about 4\%), a random uncertainty ($<$1.5\%), and a 1\% uncertainty due to the current uncertainty in the measured beam area. Note that the raw maps are calibrated in $\rm{Jy/beam}$, and we need to convert $\rm{Jy/beam}$ to be $\rm{Jy/Sr}$. 
The area of Herschel beam is simply the integral of the beam profile, where $\theta_{\operatorname{maj}}$ = $\{18.4'', 24.9'', 37.0''\}$ and $\theta_{\min }$ = $\{17.4'', 23.6'', 33.8''\}$ are the FWHM along the major and minor axes at 250, 350, and 500$\rm{\mu m}$, respectively \citep{Ott10}. The conversion from  $\rm{Jy/beam}$ to $\rm{Jy/Sr}$ involves multiplication by $\{9.065, 5.118, 2.358\} \times 10^7$ at 250, 350, and 500$\rm{\mu m}$, respectively\footnote{\tt http://herschel.esac.esa.int/hcss-doc-15.0/  SPIRE Data Reduction Guide}. 

In Figure~\ref{fig:map_cib}, we show the final Herschel-SPIRE HeLMS maps that are generated by HIPE. We can find that several bright extended stripes appear in the middle of the maps. The stripe at roughly $1^\circ < {\rm RA} < 7^\circ$ is caused by strong Galactic cirrus emission, which is confirmed by the Planck dust map \citep{Asboth18}. The stripes located at $1.2^\circ < {\rm RA} < 3.8^\circ$ and $-1^\circ < {\rm RA} < 1^\circ$ are due to Uranus and scattered light from Jupiter, respectively. We mask these regions to reduce the impact on the measurement CIB.

The entire HerMES field was observed twice in two nearly orthogonal scanning directions, and there are more than 95 \% overlapping area covered by the two scanning modes. In this work, we generated three maps (two maps with a single scan direction only, and another one with both scan directions) for each SPIRE band according to the process described above. The two maps with a single scan direction are used to analyze the noise from the instrument, and the other one is used to derive the auto and cross correlations.

\subsection{CMB lensing map from Planck}
\label{subsec:planck}
The CMB Lensing map is generated using the publicly available released Minimum-variance (MV) CMB lensing convergence $\kappa$ data from Planck Legacy Archive\footnote{\tt http://pla.esac.esa.int/pla/} \citep{Planck16XV}. The data file contains the spherical harmonic coefficients $\kappa_{\ell m}$ with $\ell _{\rm min} = 8$ and $\ell _{\rm max} = 2048$, analysis mask, the approximate noise and signal power spectrum of $\kappa$. We will discuss the detailed estimates of the theoretical auto- and cross-power spectra using the gravitational lensing potential $\phi$ in \S\ref{model}. The relationship between lensing potential $\phi$ and lensing convergence $\kappa$ can be expressed as,
\begin{equation}
\kappa_{\ell_{m}}=\frac{1}{2} \ell(\ell+1) \phi_{\ell m}.
\end{equation}

Then the gravitational lensing potential map on the sky can be defined as,
\begin{equation}
\phi(\hat{\mathbf{n}})=\sum_{\ell} \sum_{m=-\ell}^{\ell} \phi_{\ell m} Y_{\ell m}(\hat{\mathbf{n}}).
\end{equation}

\begin{figure*}[t]
\centering
\includegraphics[width=2.05\columnwidth]{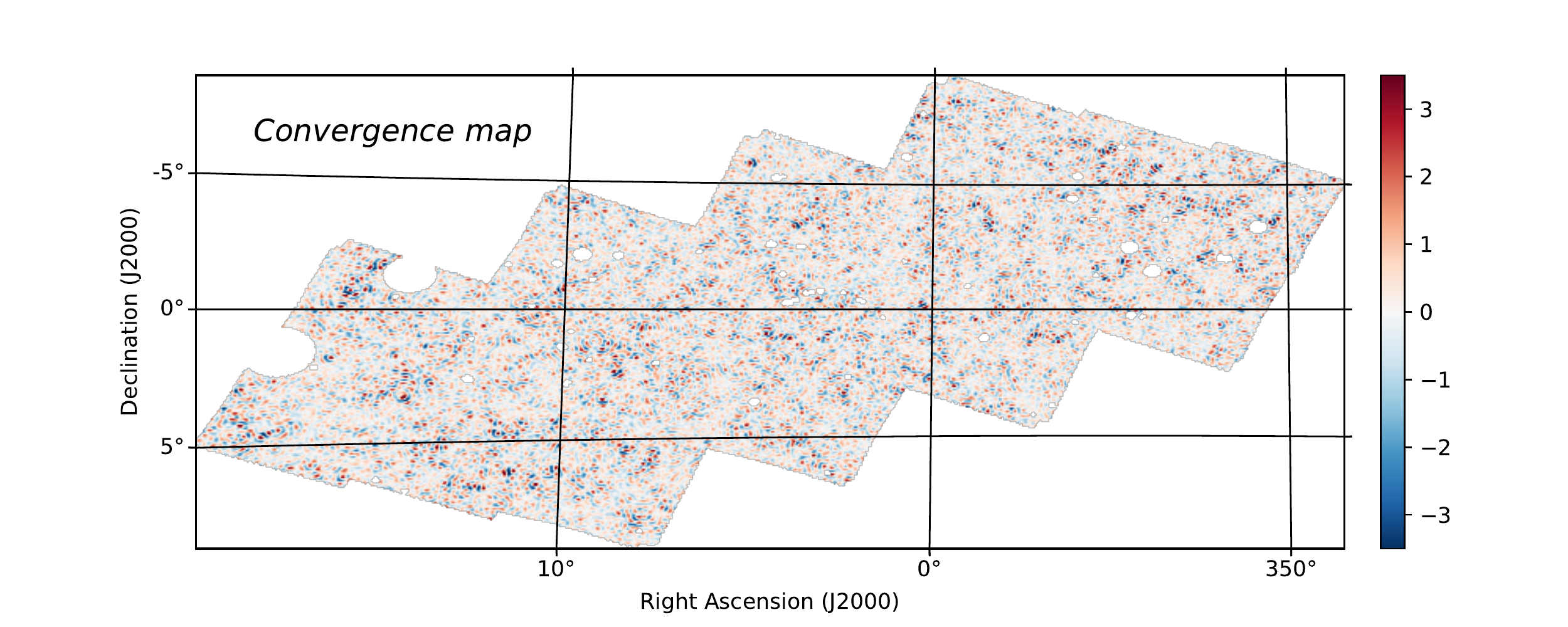}
\caption{The CMB lensing convergence $\kappa$ map in the Herschel filed.}
\label{fig:map_lensing}
\end{figure*}

We can obtain a HEALPix format lensing map with the healpix parameter $N_{\rm side} = 2048$, the pixel-scale $\sim 1.7'$ and the sky fraction $f_{\rm sky} \sim 0.67$ \citep{Gorski05}. Then we obtain the Herschel field lensing potential map and convert its format from the spherical to the planar geometry case. We show the masked lensing map in Figure~\ref{fig:map_lensing}.  Note that we smooth the Herschel map to make it the same pixel-scale ($\sim 1.7'$) as the CMB Lensing map, when we measure the cross correlation between CIB and CMB lensing.

\subsection{mask}
\label{subsec:mask}
\begin{figure*}
\centering
\includegraphics[width=2.05\columnwidth]{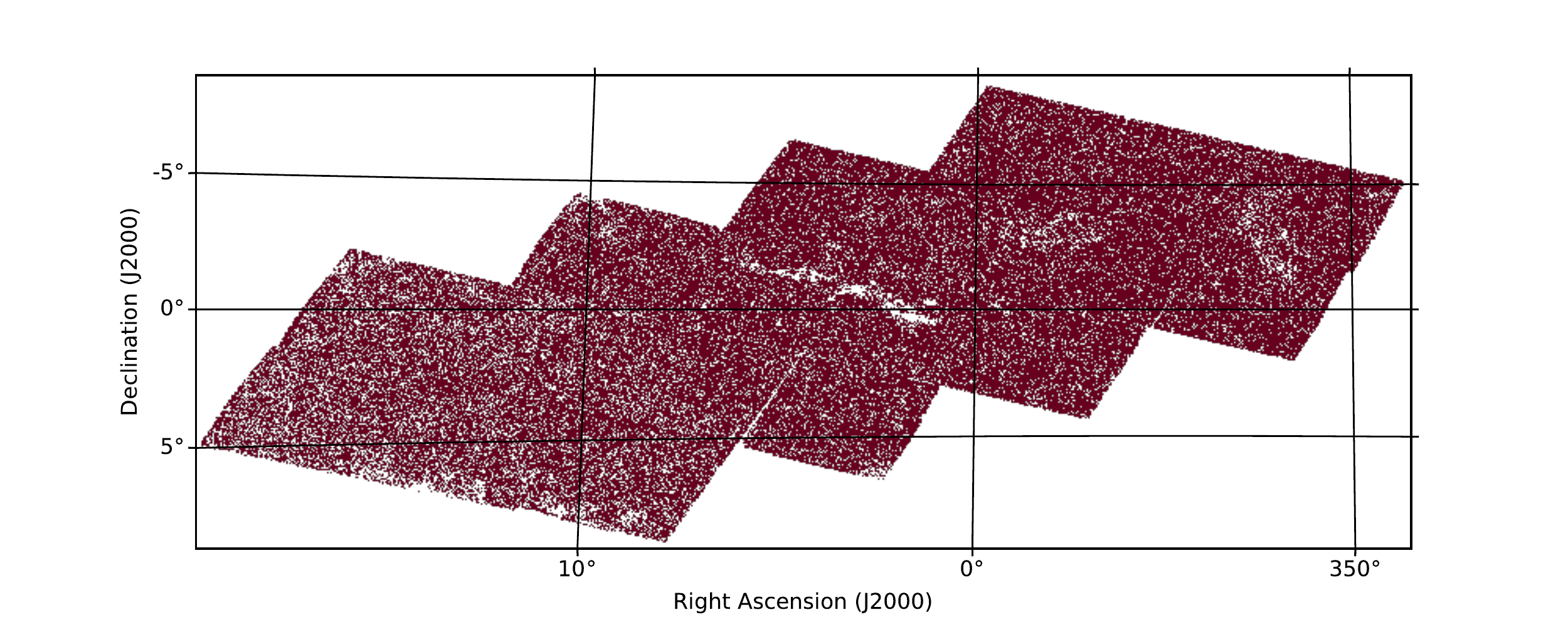}
\caption{The mask map of the Herschel maps used in this work.}
\label{fig:mask}
\end{figure*}

\begin{figure*}
\centering
\includegraphics[width=2.05\columnwidth]{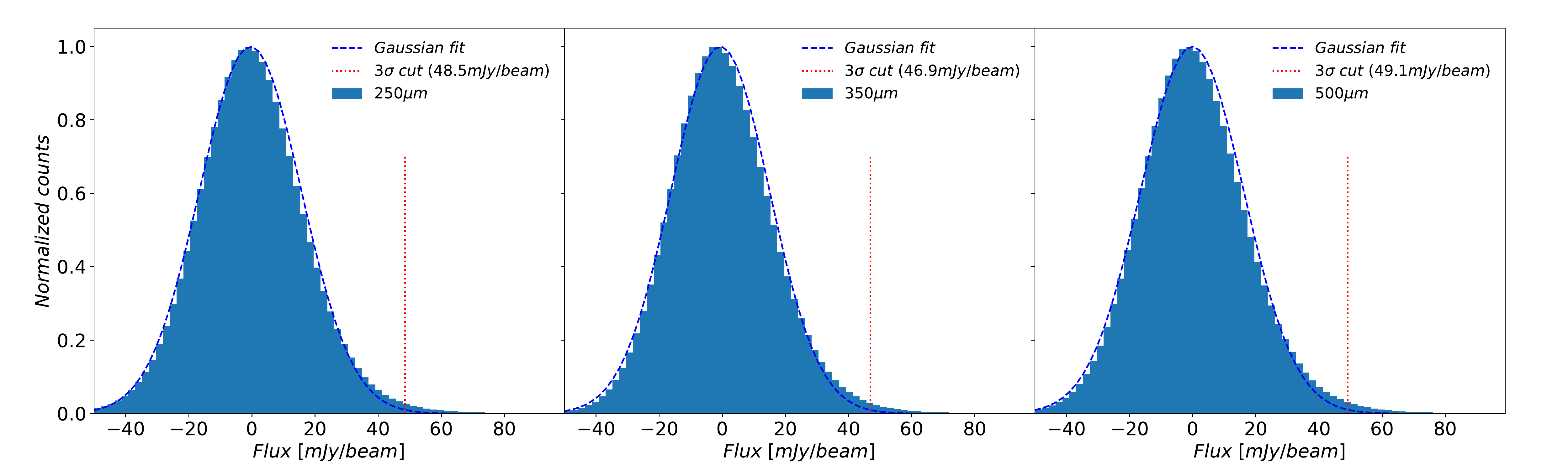}
\caption{The flux distribution of the pixel count at 250 (left panel), 350 (middle panel), and 500 $\rm{\mu m}$ (right panel), respectively. The blue dashed lines show the fitting results using the Gaussian distribution, and the red vertical dotted lines are the $3\sigma$ flux cuts.}
\label{fig:flux}
\end{figure*}

when analyzing the infrared background intensity fluctuations, we need to remove the contamination from the bright sources, especially interlopers from low redshifts. The usual way is to apply a flux cut and remove all bright pixels with flux higher than this cut. For example, \cite{Amblard11} and \cite{Thacker13} removed pixels larger than 50 $\rm{mJy/beam}$. In this work, similarly, we perform a 3$\sigma$ flux cut on the maps. In Figure~\ref{fig:flux}, we show the histograms of flux distribution of the map pixels, and find that the distribution is Gaussian-like in all three bands. The blue dashed curves show the Gaussian fitting results. Then we calculate the normalized median absolute deviation as $\sigma$ \citep{Brammer08}, and find the 3$\sigma$ flux cut  for each band. The $3\sigma$ flux cuts are 49.1, 46.9, and 48.5 $\rm{mJy/beam}$ at 250, 350, and 500 $\rm{\mu m}$, respectively, which are close to 50 $\rm{mJy/beam}$. The removed pixels includes bright stars, low-z galaxies, Galactic cirrus, Uranus, and scattered light from Jupiter. We also remove the areas that do not contain measured data.
Finally, we combine the masks of all bands into a final mask map. We show the final CIB mask map in Figure~\ref{fig:mask}. This mask removes about 20\% pixels in the scanned area.

In the Planck lensing data, the mask obtain from file $mask.fits$ of $COM\_\ Lensing\_\ 2048\_\ R2.00.tar$ in Planck Legacy Archive. The masked source (white holes) is easy to find in Figure~\ref{fig:map_lensing}. By combing the mask used in the CIB auto-correlation analysis shown in Figure~\ref{fig:mask}, we obtain the mask map used in the calculation of the cross-correlation of CIB and CMB lensing.

\section{power spectrum measurements}
\label{sub:spectrum}
In this work, we calculate the auto- and cross-power spectra following the method given by \cite{Cooray12}, and the $C_{\ell}$ is then given by
\begin{equation}
 C_{\ell_{i}}=\frac{\sum_{\ell_{1}}^{\ell_{2}} w\left(\ell_{x}, \ell_{y}\right) \widetilde{\mathcal{M}}_{X}\left(\ell_{x}, \ell_{y}\right) \widetilde{\mathcal{M}}_{\mathrm{Y}}^{*}\left(\ell_{x}, \ell_{y}\right)}{\sum_{\ell_{1}}^{\ell_{2}} w\left(\ell_{x}, \ell_{y}\right)}.
\end{equation}

Here $\ell_i$ id the $\ell$-mode of the $i$-th bin between $\ell_1$  and $\ell_2$, and $\ell_i=(\ell_1+\ell_2)/2$, where $\ell_1^2 < \ell_x^2 +\ell_y^2 <\ell_2^2$, $w\left(\ell_{x}, \ell_{y}\right)$ is a weighting function in Fourier space, which is non-zero for modes used in the analysis, and zero for modes that are discarded. The weighting function is obtained by calculating the Fourier transform of the mask. $\widetilde{\mathcal{M}}\left(\ell_{x}, \ell_{y}\right)$ is the two-dimensional Fourier transform of the observed map. For cross-power spectrum, $X$ and $Y$ represent two different maps, and $X=Y$ for the auto-power spectrum.

The raw $\widetilde{C}_{\ell}$ directly measured from the masked observed maps is a pseudo angular power spectrum. It is affected by map-making process, mask, instrumental beam and noise. The relation between the final and raw power spectrum is described as,
 
\begin{equation}
\widetilde{C}_{\ell}=B^{2}_{\ell}T_{\ell} M_{\ell \ell^{\prime}} C_{\ell^{\prime}}+N_{\ell},
\end{equation}
where $C_{\ell^{\prime}}$ is the true angular power spectrum from full sky, $B_{\ell}$ is the beam function, $T_{\ell}$ is the map-making transfer function, $M_{\ell \ell^{\prime}}$ is the mode coupling matrix, and $N_{\ell}$ is the instrumental noise.

\subsection{Noise Model}
\label{Noise}
We generate the instrumental noise using a simple noise model, which assumes that the noise power spectrum should be smooth as a function of $\ell$, and it can be written as follows \citep{Amblard11,Thacker13},
\begin{equation}
N_{\ell}=N_{0}\left[\left(\frac{\ell_{0}}{\ell}\right)^{2}+1\right],
\end{equation}
where the noise is almost white-noise ($N_{\ell}\to$constant) at small scales (large $\ell$), and shown as the 1/f-type ($N_{\ell}\propto \ell^{-2}$) at large angular scales (small $\ell$). 

\begin{figure}
\centering
\includegraphics[width=1.0\columnwidth]{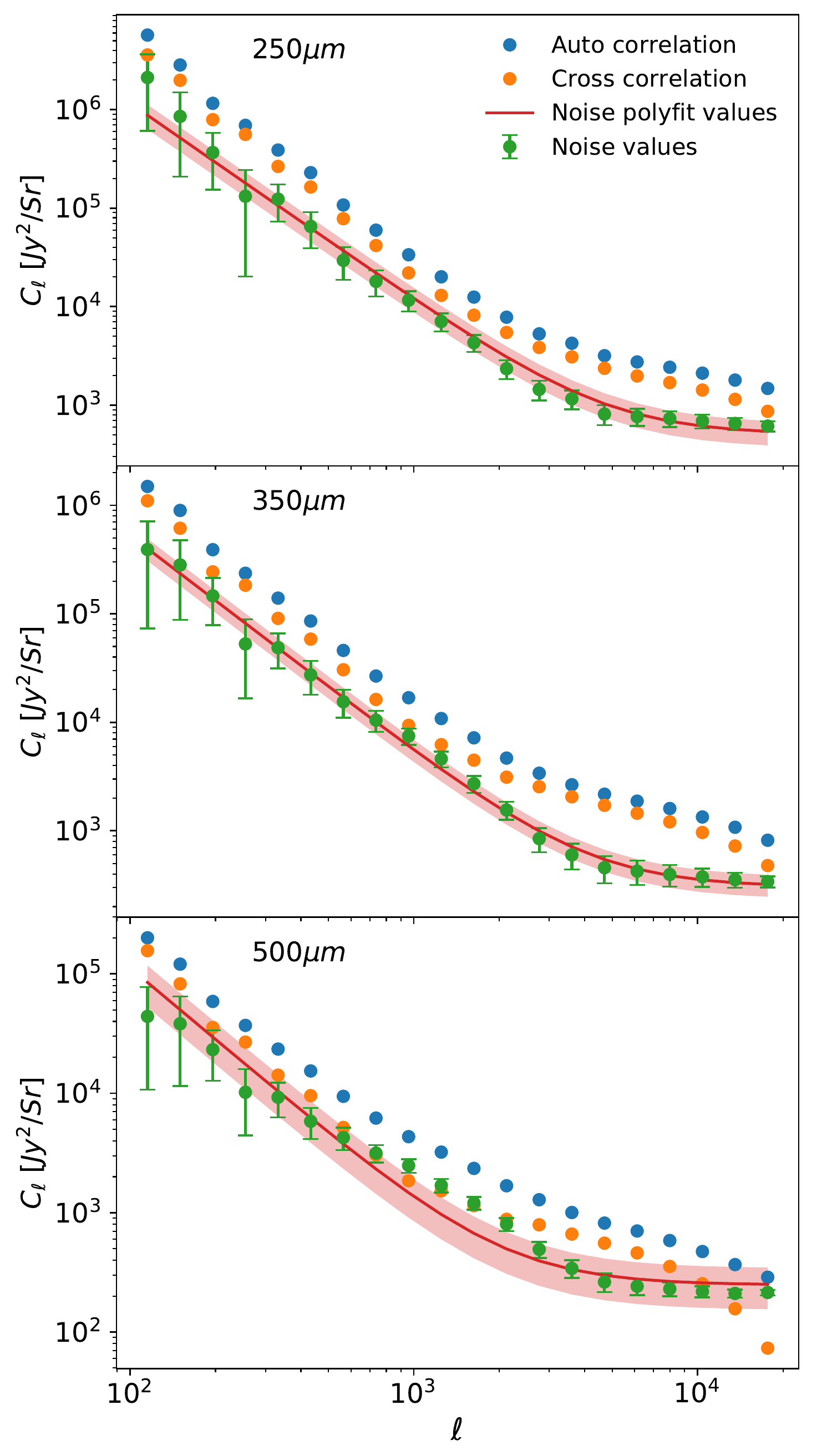}
\caption{The instrumental noise power spectrum at 250 (top panel), 350 (middle panel), and 500 $\rm{\mu m}$ (bottom panel), respectively. The blue dots show the auto-power spectrum from the final Herschel-SPIRE HeLMS maps which is shown in Figure~\ref{fig:map_cib}, and it is composed of the sky signal and instrumental noise. The orange dots show the cross-power spectrum of two maps with nearly orthogonal scan directions. The green dots with error bars show the instrumental noise power spectrum, which are obtained by calculating the difference between of the auto- and cross-power spectrum. The solid red lines show the best-fit noise curves, and shaded regions show the $1\sigma$ uncertainties}
\label{fig:noise}
\end{figure}

The instrumental noise bias in the power spectrum can be avoided by cross-correlating two different imaging maps in the same field. Since almost each area of the HeLMS field is observed twice, we can obtain two sets of maps.  Each map only contains a single scan direction, and the scan directions are nearly orthogonal for the two maps in the same survey area. We also generate the N-hits maps which give the number of bolometer hits per sky pixel. We find that almost all of pixels are covered by both directional scans. Although the two maps used here have nearly orthogonal scan directions, we have made sure that they are cross-linked for the pixels visited by both directional scans. A few uncross-linked pixels are masked in our analysis. We derive the power spectrum of instrumental noise by calculating the difference between of the auto- and cross-power spectrum in the same field. Since the auto power spectrum contains both signal and noise while the cross power only includes the former, the subtraction between them can extract the noise term. We use the least square method to fit the noise curves, and get the value of noise parameters $N_0$ and $\ell_0$. We show the power spectrum of the instrumental noise with errors and the fitting results in Figure~\ref{fig:noise}. The fitting values of $N_{0}$ and $\ell_0$ at 250, 350, and 500 $\rm{\mu m}$ bands are shown in Table~\ref{tab:n0l0}. Note that the observing mode of the HeLMS field is using the $Large$ $Map$ mode (fast scan mode) with a scan rate of $60''\ \rm s^{-1}$, compared to the $Parallel$ Mode (slow scan mode) with $20''\ \rm s^{-1}$ used in other studies  \citep[e.g.][]{Thacker13}. The larger survey area and faster scan rate can lead to some differences of the noise power spectrum compared to other works.

\begin{table}
\centering
\caption{The  fitting result of noise model parameters $N_{0}$ and $\ell_0$ .}
\begin{tabular}{c c c c }
\hline\hline
 \rule[-2mm]{0mm}{6mm}
Parameter & 250$\rm{\mu m}$ & 350$\rm{\mu m}$ & 500$\rm{\mu m}$   \\
 \hline
 $N_{0} $ & 504 & 303  & 247 \\
 $\ell_0$ & 4805 & 4180 & 2128 \\
 \hline\end{tabular}
\label{tab:n0l0}
\end{table}

\subsection{Beam Correction}
Due to the detector resolution limits, there is a non-negligible drop in the raw CIB power spectrum, especially at small scales $\ell\gtrsim10^4$. We use a beam function $B_{\ell}$ to correct the raw power by $C_{\ell}=\widetilde{C}_{\ell} / B_{\ell}^2$. 
Since the Herschel beam is not Gaussian, if using a 2D-Gaussian beam approximation, they are about 15\%, 10\%, and 10$\%$ smaller than the measured Herschel beam areas at 250 $\rm \mu m$, 350 $\rm \mu m$, and 500 $\rm \mu m$, respectively. So we calculate the beam function according to \cite{Amblard11}, the beam function is obtained directly from the beam profiles, it can be expressed as,

\begin{equation}
B_{\ell}^2=\frac{C_{\ell}^{\rm beam}}{C_{\ell}^{\rm point}}.
\label{eq:beam}
\end{equation}
Here $C_{\ell}^{\rm point}$ is the power spectrum of a point source where all the light lies in one pixel, $C_{\ell}^{\rm beam}$ is the power spectrum of Herschel beam profiles, which have been measured by studying Neptune\footnote{\tt http://hedam.oamp.fr/HerMES/index/dr4}.
The mean FWHM values are $17.9''$, $24.2''$, and $35.4''$, and the mean ellipticities are $5.1\%$, $5.4\%$, and $8.7\%$ for the  maps at 250, 350, and 500 $\rm{\mu m}$, respectively \citep{Griffin10}. The systematic variations are about $5\%$. Eq.~(\ref{eq:beam}) is also applied to the cross-correlation between maps $X$ and $Y$ with $B_{\ell}=\sqrt{b_X b_Y}$, where $b_X$ and $b_Y$ are the beam functions of auto-correlation of $X$ and $Y$, respectively.

\subsection{Mode Coupling Correction}

In \S\ref{subsec:mask}, we mask the bright sources and the areas that do not contain measured data. This process leads to some fictitious information when performing Fourier transform for converting into the final power spectrum. An easy way to correct the mask is to divide the raw power spectrum by the masked sky fraction $f_{\rm sky}$. Another method is more complicated, which utilizes the mode coupling matrix \citep[e.g.][]{Cooray12,Thacker15}. The mask can break the large $\ell$ modes into small ones, so that the profile of the power spectrum will be changed by the mask effect. Here it is more accurate using the mode coupling matrix. As a fast and accurate method, MASTER is widely used in calculation of mode coupling matrix. \cite{Cooray12} and \cite{Zemcov14} extended the MASTER method from analytical calculation to simulation. 

\begin{figure}
\centering
\includegraphics[width=1.0\columnwidth]{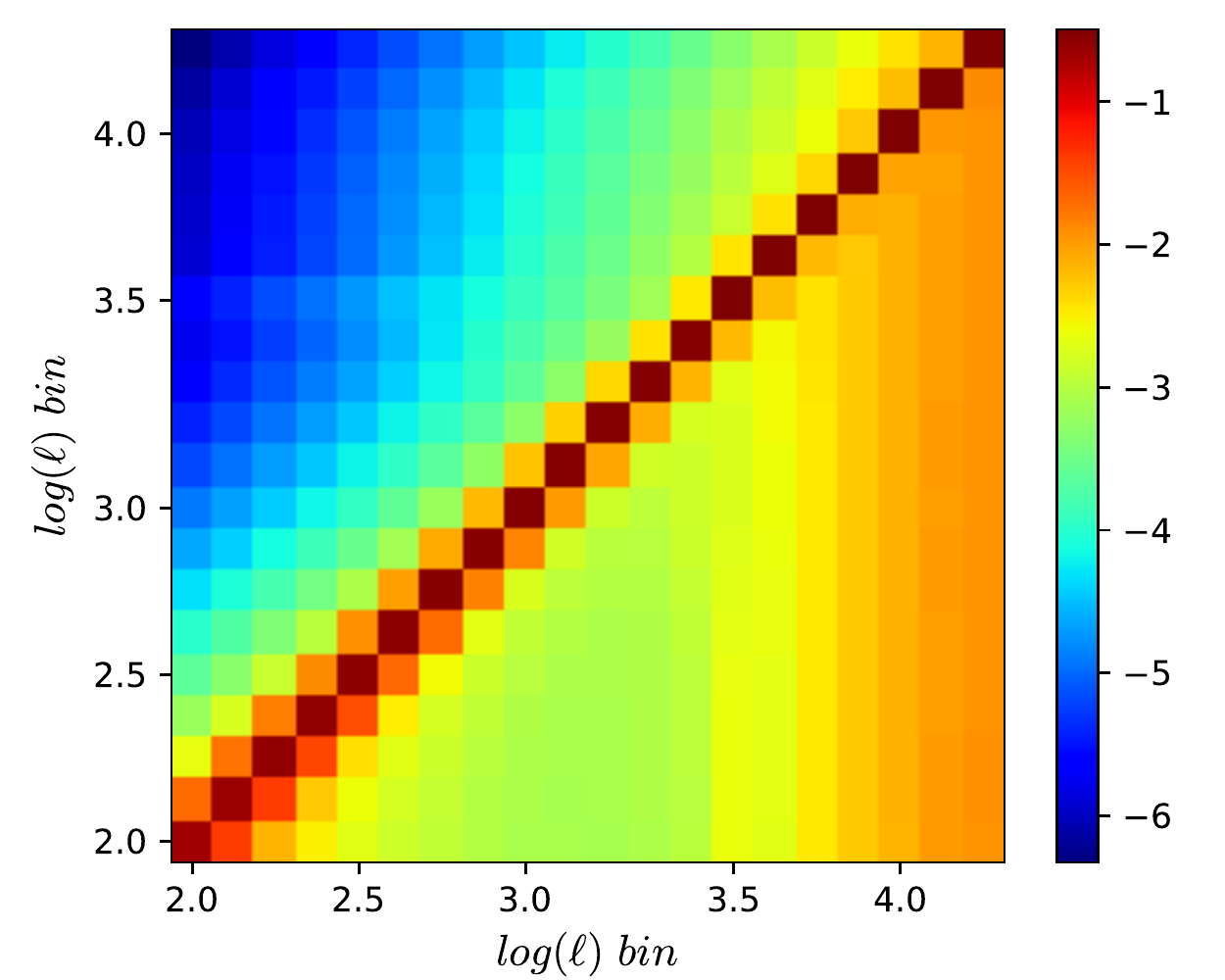}
\caption{An example of mode coupling matrix $M_{\ell\ell^{\prime}}$ for CIB auto-correlation with $\ell$ mode equally divided into 20 bins from 100 to 20000 (left to right) in logarithmic scale, and the coordinates of the color bar is also in logarithmic scale. we can find that the effect between adjacent bins is greater, and large $\ell$ modes (small scale) has a greater effect than small $\ell$ modes (large scale).  }
\label{fig:mll}
\end{figure}

In this work, the mode coupling matrix $M_{\ell \ell^{\prime}}$ is generated by a three-step process in each bands. First, we generate 100 simulation realizations of maps from a pure tone power spectrum, where $C_{\ell^{\prime}} = 1$ if $\ell$ in $i$-th bin, otherwise $C_{\ell^{\prime}} = 0$. Next, we mask these realizations using the real space mask that is discussed in \S\ref{subsec:mask}, and then calculate the raw power spectra $\widetilde{C}_{\ell}$ of these masked maps. So the $i$-th row of mode coupling matrix $M_{\ell \ell^{\prime}}$ can be expressed as $\left\langle{\widetilde{C}}_{\ell}\right\rangle$. Finally, We repeat the above process for all $\ell$ bins to obtain the $M_{\ell \ell^{\prime}}$. The results of the mode-coupling matrix at 250 $\rm{\mu m}$ shown in Figure~\ref{fig:mll}. We use the inverse of $M_{\ell \ell^{\prime}}$ to estimate the unmasked power spectrum by $C_{\ell^{\prime}}= M_{\ell \ell^{\prime}}^{-1}\widetilde{C}_{\ell}$.

\subsection{Transfer Function}
\label{subsec:trans_func}

Due to finite detectors, we can only observe large sky filed by scanning, and obtain the data by merging all scan time streams. Thereby, the detector arrays, scanning method and pipeline process result in an imperfect representation of the sky. In this work, we estimate the true power spectrum of the sky by $C_{\ell}= \widetilde{C}_{\ell} /T_{\ell}$. 

We calculate the transfer function $T_{\ell}$ by using a simulation method, that can be described as a three-step process. First, we generate randomly 100 simulated Gaussian realizations of the sky, and obtain the time streams with the same scan path as the real observation. Next, we subtract the median value in each time stream, and merge the processed time streams into final maps. Finally, the map-making transfer function can be described by $T_{\ell} =\left\langle \widetilde{C}_{\ell}/C_{\ell}\right\rangle$, where $C_{\ell}$ is the input power spectrum of the simulated sky and $\widetilde{C}_{\ell}$ is the power spectrum of the simulation map. We also repeated all steps above for different input power spectra. The result show that map-making transfer function is independent of the input power spectrum. 

\begin{figure}
\centering
\includegraphics[width=1.07\columnwidth]{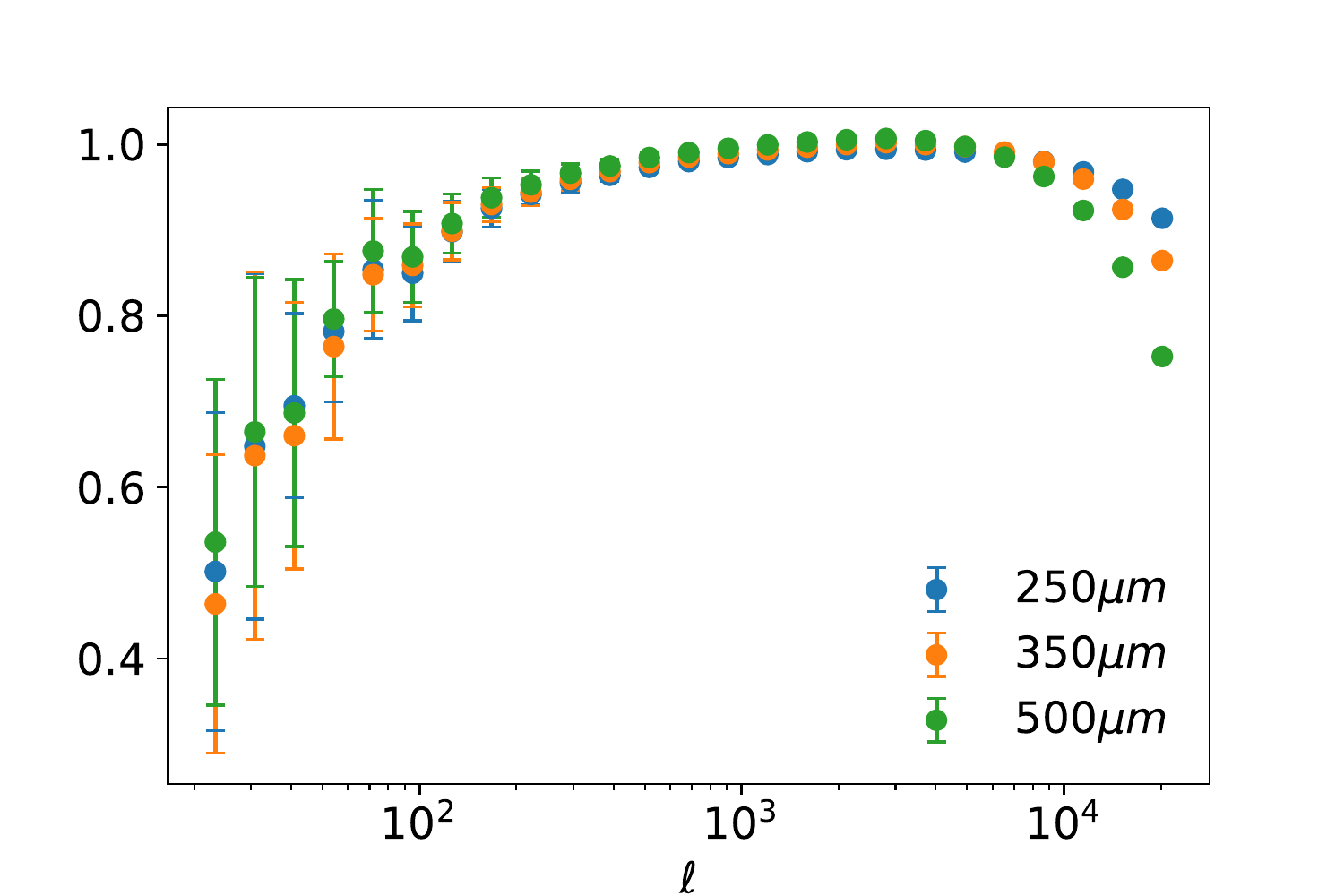}
\caption{The map-making transfer function at 250 (blue points), 350 (orange points), and 500 $\rm{\mu m}$ (green points), respectively. These data points and uncertainties are calculated from 100 simulations of Gaussian realizations. The values are almost same at large angular scales (small $\ell$ modes) for the three bands, and the attenuation is mainly caused by removing the medium value in individual time stream. On the other hand, the values of $T_{\ell}$ are different at small angular scales (lager $\ell$ modes), and the suppression feature may be due to the scan pattern or pixel window function effect.}
\label{fig:tran}
\end{figure}

The map-making transfer functions at 250, 350, and 500 $\rm{\mu m}$ are show in Figure~\ref{fig:tran}. We can see that the $T_{\ell}$ is almost the same at large angular scales (small $\ell$ modes) for the three bands, and the attenuation feature is mainly due to removing the medium value in individual time stream. At small scales, $T_{\ell}$ are different for the three bands, and the suppression may be caused by the scan pattern that leaves stripes in the data or the effect of pixel window function for different pixel sizes \citep{Viero13,Thacker15}. We calculate the standard deviation of the 100 simulations as the uncertainties of the map-making transfer function. For the lensing map, we use a simple assumption that $T_{\ell} = 1$.

\subsection{Statistical error of Power Spectrum}

The statistical error $\delta C_{\ell}$ of the auto-power spectrum can be obtain by \citep{Cooray12,Zemcov14,Mitchell15,Thacker15},

\begin{equation}
\delta C_{\ell}=\sqrt{\frac{2}{f_{\rm sky}(2 \ell+1) \Delta \ell}}\left(C_{\ell}+N_{\ell}\right),
\end{equation}
where $f_{\rm sky}$ is the fraction of the unmasked areas of all sky, $ \Delta \ell$ is the width of the bin, $N_{\ell}$ is the instrument noise that described in \S\ref{Noise}.

For the cross-correlation, the statistical error becomes \citep{Planck11XVIII,Zemcov14,Thacker15},
\begin{equation}
\delta C_{\ell}^{XY}=\sqrt{\frac{\left(C_{\ell}^{X}+N_{\ell}^{X}\right)\left(C_{\ell}^{Y}+N_{\ell}^{Y}\right)+\left(C_{\ell}^{XY}\right)^{2}}{f_{\rm sky}(2 \ell+1) \Delta \ell}},
\end{equation}
where $C_{\ell}^{X}$ and $C_{\ell}^{Y}$ are the auto-correlation power spectrum, and $N_{\ell}^{X}$ and $N_{\ell}^{Y}$ are the noise from map $X$ and $Y$, respectively. $C_{\ell}^{XY}$ is the cross-power spectrum.

We can also estimate the signal to noise ratio (SNR) of the power spectra for both auto and cross correlations, and it is given by
\begin{equation}
{\rm SNR}=\sqrt{\sum_{\ell\ {\rm bin}} \left[\frac{C_{\ell}^{XY}}{\delta C_{\ell}^{XY}}\right]^2}.
\end{equation}

\begin{figure*}
\centering
\includegraphics[width=1.0\columnwidth]{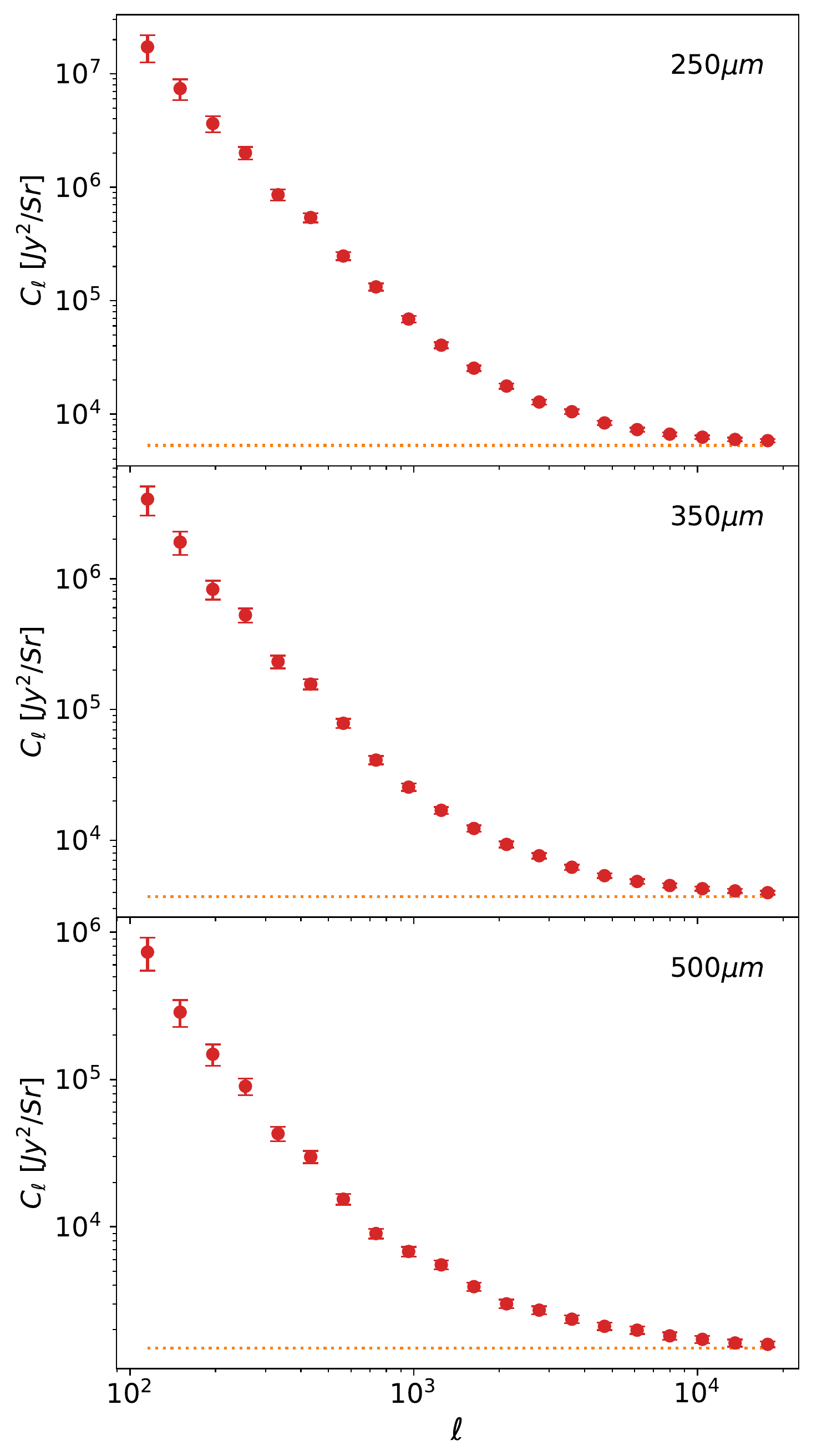}
\includegraphics[width=1.0\columnwidth]{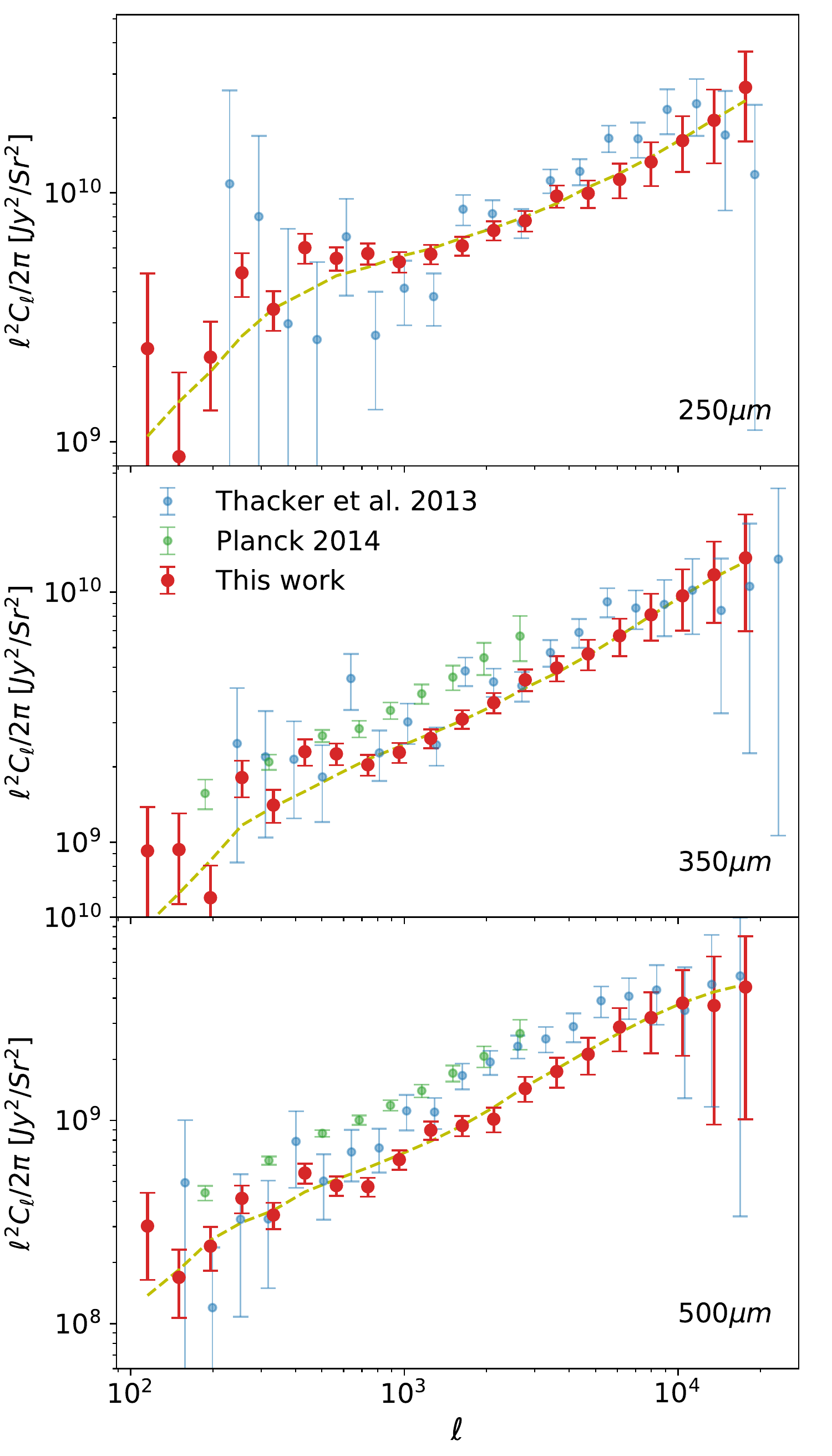}
\caption{$Left:$ The red data points are the final angular power spectrum $C_{\ell}$ at 250 (top panel), 350 (middle panel), and 500 $\rm{\mu m}$ (bottom panel), respectively. The errors are obtained by calculating the square root of instrument noise and beam, map-making transfer function and the cosmic variance. The orange dotted lines are the best-fit shot noise power spectra fitted by the MCMC method. $Right:$ The CIB power spectra without shot noise and DGL components are shown in the right panels as red data points. The blue and green data points represent the results given by \cite{Thacker13} and \cite{Planck14XXX}, respectively. The yellow dashed curves are the best-fitting CIB power spectra from the model.}
\label{fig:auto} 
\end{figure*}

\subsection{Final Power Spectra}
\label{subsec:final_power}

We obtain the Herschel-SPIRE HeLMS field final auto- and cross-power spectra using the above estimates and corrections, and show them in Figure~\ref{fig:auto} and Figure~\ref{fig:cross}. The values are shown in Table~\ref{tab:auto} and Table~\ref{tab:cross} in appendix, respectively. 

In the left panel of Figure~\ref{fig:auto}, we can find the CIB auto-power spectrum is almost a constant at small scales (large $\ell$), where it is dominated by the shot noise (the orange dotted lines). The amplitude of the shot noise  depends on the flux cut of masked sources \citep{Viero13}. We use a $3\sigma$ flux cut here, and fit the $C_{\ell}$ measurements based on MCMC method at small angular scales (lager $\ell$ modes), then we obtain the shot noise and their uncertainties. We find that they are  $5.33^{+0.14}_{-0.15}\times10^3$, $3.70^{+0.09}_{-0.09}\times10^3$, and $1.50^{+0.06}_{-0.06}\times10^3$ $\rm{Jy^2/Sr}$ at 250, 350, and 500 $\rm{\mu m}$, respectively. 

\begin{figure}
\centering
\includegraphics[width=1.0\columnwidth]{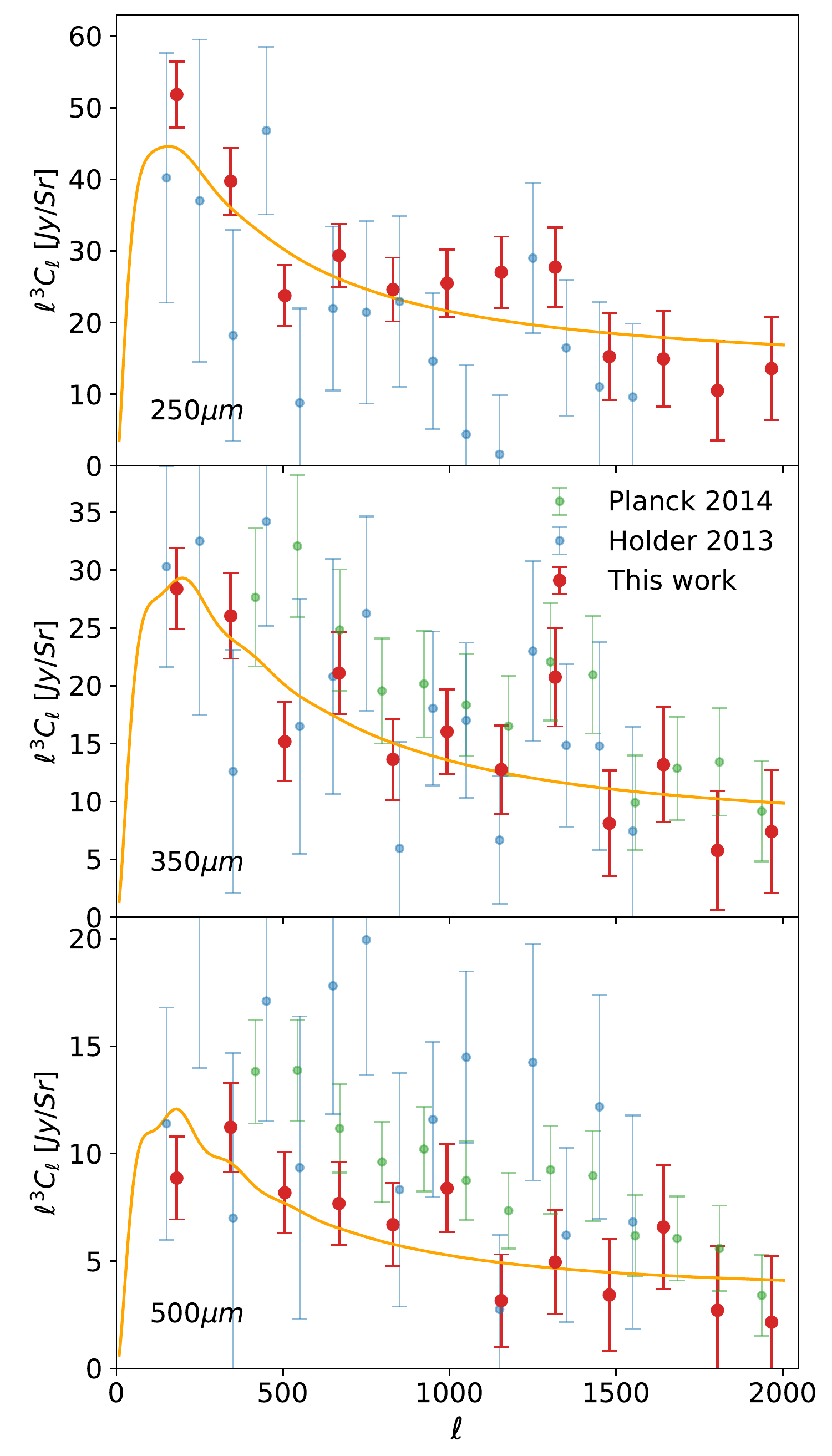}
\caption{The cross power spectra of the CIB and CMB lensing at 250 (top panel), 350 (middle panel), and 500 $\rm{\mu m}$ (bottom panel) are shown in red data points. The blue and green points represent the results given by \cite{Holder13} and \cite{Planck14XVIII}, respectively. The orange curves are the best-fitting cross power spectra given by the MCMC method.}
\label{fig:cross} 
\end{figure}

In the right panel of Figure~\ref{fig:auto}, we show the final power spectrum subtracting the shot noise and the diffused Galactic light (DGL) components, and compare them to the power spectra given by \cite{Thacker13} (GAMA fields of H-ATLAS) and Planck team \citep{Planck14XXX}. As can be seen, our results (the red data points) are generally consistent with \cite{Thacker13} (blue data points) and \citep{Planck14XXX} (green data points). In \cite{Thacker13}, they also explore Herschel data in Herschel-ATLAS GAMA-15 field, and as can be seen, we have smaller error bars and can explore larger scales at $\ell\simeq 100$, since the area of HeLMS field we use is much larger than GAMA-15 field. The SNRs we obtain are 15.9, 15.7 and 15.3 at 250, 350, and 500 $\rm \mu m$ respectively. Compared with the results of \cite{Thacker13}, the SNRs are improved by a factor of $\sim$2 averagely for the three bands. Besides, we can provides better measurements on the DGL component, which is dominant at large scales, and we will discuss it in details in the next section. On the other hand, comparing to \citep{Planck14XXX}, we can obtain comparable results with similar error bars over larger scale range from $\ell=100$ to 20000.

In the Figure~\ref{fig:cross}, the measured cross-power spectra of the CIB and CMB lensing at 250, 350, and 500 $\rm \mu$m are shown. We also compare the cross-power spectra to the results of \cite{Holder13} (blue data points) and \cite{Planck14XVIII} (green data points).
Note that \cite{Planck14XVIII} provided the measurements of the cross-correlation spectra of CIB anisotropies and CMB lensing at 545 and 857 GHz, and we rescale the Planck 545 GHz data by a factor of 1.22 to match to the Herschel data at 500 $\rm \mu m$ \citep{Hanson13}, since Planck measured the data at 550 $\rm \mu m$ for 545 GHz.  \cite{Holder13} presented the measurements of cross-correlation of gravitational convergence and Herschel-SPIRE maps covering 90 deg$^2$ at 250, 350, and 500 $\rm \mu m$. They used a CMB map obtained by the South Pole Telescope at 150 GHz to construct the gravitational convergence map \citep{Holder13}. We can find that our results are in good agreements with both \cite{Planck14XVIII} and \cite{Holder13}, and we have relatively higher accuracy measurements with smaller error bars, especially at large scales. The SNRs are 7.5, 7.0, and 6.2 at 250, 350, and 500 $\rm \mu$m, respectively, in our analysis. It is about 10-20\% better than the results of \cite{Planck14XVIII}, and a factor of 2$\sim$3 better than \cite{Holder13}.

\section{theoretical model and analysis}
\label{model}

\begin{figure*}
\centering
\includegraphics[width=0.68\columnwidth]{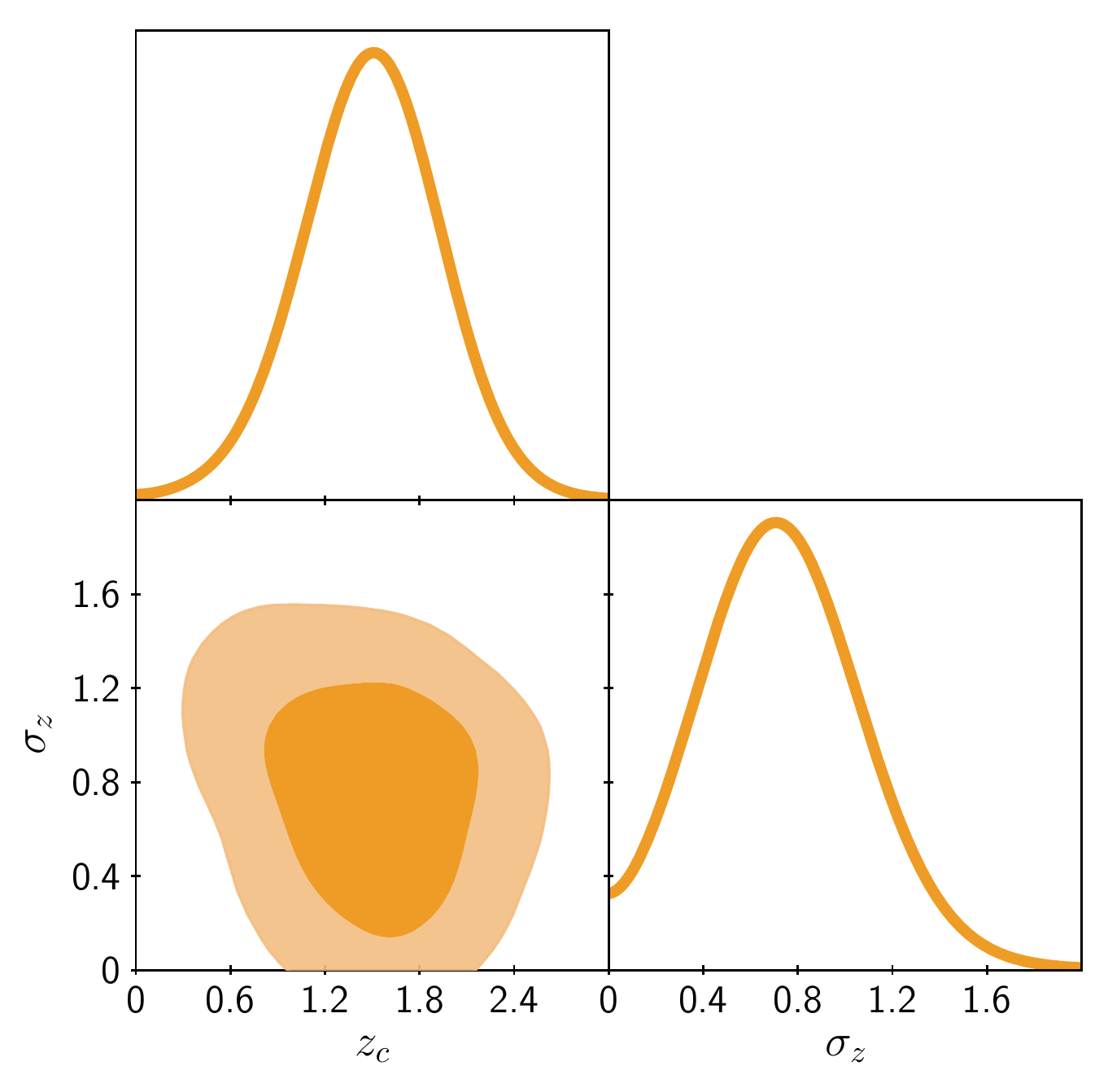}
\includegraphics[width=0.68\columnwidth]{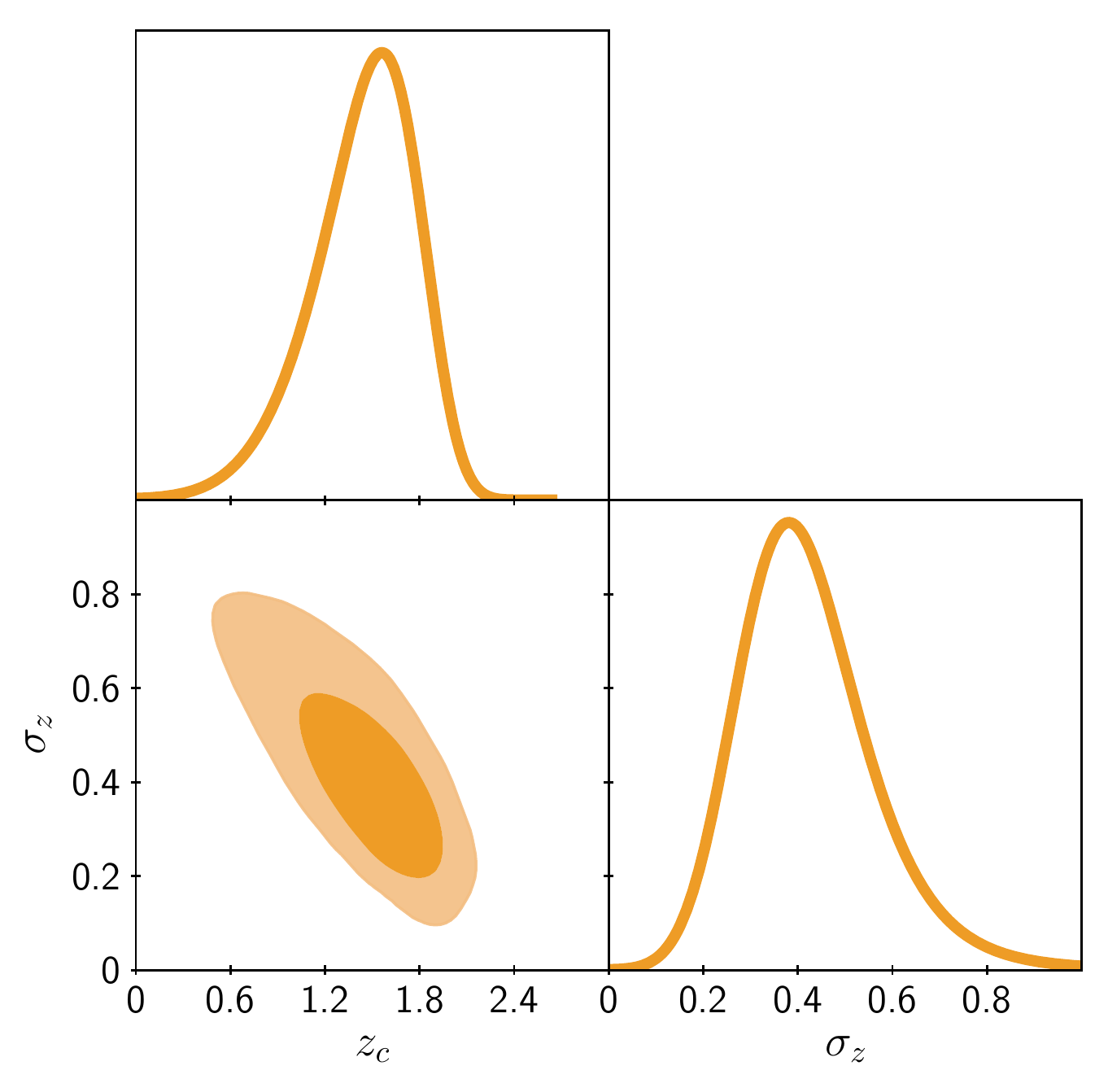}
\includegraphics[width=0.68\columnwidth]{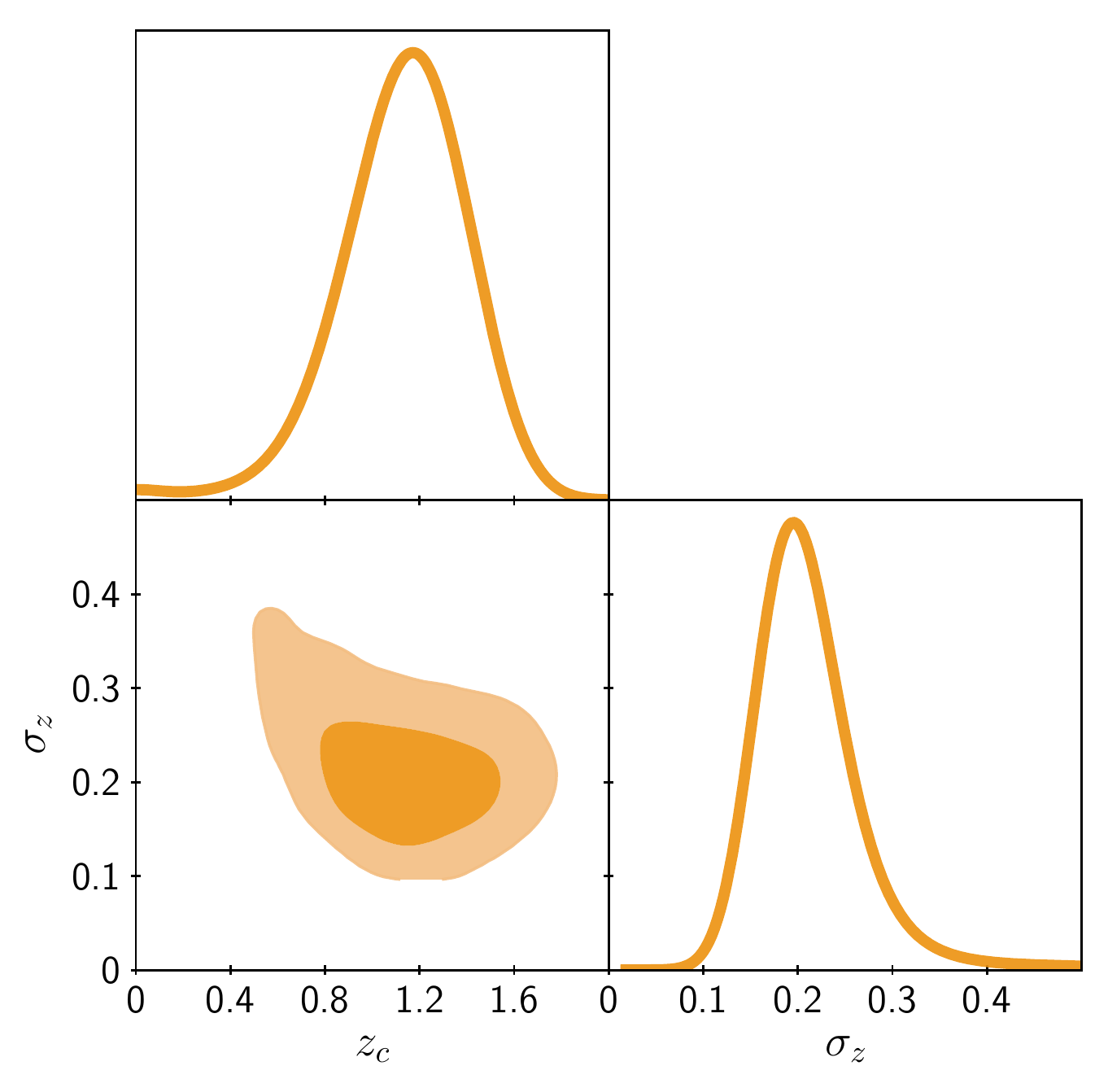}
\caption{ The contour maps with 1$\sigma$ and $2\sigma$ C.L. of $z_c$ and $\sigma_z$ at 250 (left panel), 350 (middle panel), and 500 $\rm{\mu m}$ (right panel), respectively. The solid lines are the 1-D PDFs of the parameters. }
\label{fig:fit} 
\end{figure*}

\subsection{model}
The source at the last scattering surface is lensed by the gravitational potential of all mater from us to the last scattering surface, so the lensed temperature anisotropies $\Theta(\hat{\mathbf{n}})$ can be as a remapping of the primary temperature anisotropies $\widetilde{\Theta}(\hat{\mathbf{n}})$ by a two-dimensional vector field:
\begin{equation}
\Theta(\hat{\mathbf{n}})=\widetilde{\Theta}(\hat{\mathbf{n}}+\nabla \phi),
\end{equation}
where $\phi$ is the CMB lensing potential which is related to three-dimensional gravitational potential $\Psi\left(\chi \hat{\mathbf{n}} ; z\right)$ \citep{Hu01,Lewis06}, we have:
\begin{equation}
\phi(\hat{\mathbf{n}}) = -2 \int_{0}^{z_{*}}dz \frac{d\chi}{dz} \frac{f_{\Omega_K}(\chi_{*}-\chi)}{f_{\Omega_K}(\chi_{*})f_{\Omega_K}(\chi)} \Psi\left(\chi \hat{\mathbf{n}} ; z\right),
\end{equation}
where $z_*$ and $\chi_*$ respectively denote the redshift and comoving distance of last scattering surface, and the comoving angular diameter distance $f_{\Omega_K}(\chi) = \chi$ in flat universe. The comoving distance along the sight $\chi(z)$ is defined by

\begin{equation}
\chi(z) = D_H\int_{0}^{z} \frac{H_0}{H(z')}dz',
\end{equation}
where $H_0 = 100h\ \rm{km\ s^{-1}\ Mpc^{-1}}$ is the Hubble parameter today, and $D_H$ is the Hubble distance today, where $D_H = c / H_0$. 

We use the far-infrared background model which is studied by \cite{Knox01}, and the CIB mean intensity at frequency $\nu$ is related to the CIB mean emissivity $\overline{j}_{\nu}(z)$ via

\begin{equation}
I_{\nu}(\hat{\mathbf{n}})=\int_{0}^{z_*} dz \frac{d \chi}{dz} a \overline{j}_{\nu}(z)\left[1+\frac{\delta j_{\nu}(\chi \hat{\mathbf{n}},z)}{\overline{j}_{\nu}(z)}\right],
\end{equation}
where $a$ is the scale factor. We decompose the lensing potential and the CIB mean intensity into spherical harmonic coefficients by

\begin{equation}
\begin{split}
\phi_{\ell m} &=\int d \hat{\mathbf{n}} \phi(\hat{\mathbf{n}}) Y_{\ell m}^*(\hat{\mathbf{n}}),\\
t_{\nu , \ell m} &=\int d \hat{\mathbf{n}} I_{\nu}(\hat{\mathbf{n}}) Y_{\ell m}^*(\hat{\mathbf{n}}).
\end{split}
\end{equation}\\

The angular power spectrum are defined by

\begin{equation}
\left\langle X_{\ell m} Y_{\ell^{\prime} m^{\prime}}^{*}\right\rangle =\delta_{\ell \ell^{\prime}} \delta_{m m^{\prime}} C_{\ell}^{XY},
\end{equation}\\
where X and Y are the CIB at observed frequency $\nu$ or the CMB lensing potential $\phi$.

At small angular scales ($\ell \ge 100$), we can calculate the power spectra using the Limber approximation \citep{Limber53}, we have

\begin{equation}
{C}^{\rm XY}_{\ell} = \int_{0}^{z_*}dz\frac{d\chi}{dz}W^X(z)W^Y(z)P(k=\ell/\chi,z),
\end{equation}
where $P(k,z)$ is the matter power spectrum at redshift $z$. We calculate $P(k,z)$ using the nonlinear Halofit model by the Cosmic Linear Anisotropy Solving System (CLASS) \citep{Blas11}. $W(z)$ is the kernel function that indicates the weight at redshift $z$ for each of the above signals

\begin{equation}
\begin{split}
W^{\nu}(z) &= b \frac{a \bar{j}_{\nu}(z)}{\chi(z)},\\
W^{\phi}(z) &=-\frac{3 \Omega_m}{a} \left(\frac{H_0}{\ell c}\right)^2 \left[\frac{\chi_* - \chi(z)}{\chi_*}\right],
\end{split}
\end{equation}\\
where $b$ is the mean dusty star-forming galaxy bias, $\Omega_m$ is the value of the current total matter density and $\bar{j}_\nu(z)$ is the mean emissivity of CIB at frequency $\nu$, we write the CIB mean emissivity as \citep{Hall10,Planck14XVIII},

\begin{equation}
\bar{j}_{\nu}(z) = A_j a \chi^{2}(z) \exp \left[-\frac{\left(z-z_{c}\right)^{2}}{2 \sigma_{z}^{2}}\right]f_{\nu(1+z)},
\label{eq:jv}
\end{equation}
where $A_j$ is the amplitude parameter of $\bar{j}_\nu(z)$, which is degenerate with galaxy bias $b$. So we use a parameter $A$ instead of the combined values of $b\times A_j$.

In this model, we assume that all galaxies can be described by a graybody spectrum
\begin{equation}
f_{\nu} \propto \nu^{\beta} B_{\nu}\left(\nu,T_{d}\right),
\end{equation}
where $B_{\nu}\left(\nu,T \right)$ is the spectral radiance of a blackbody for frequency $\nu$ at temperature $T$, $\beta$ is the emissivity spectral index of thermal graybody dust, $T_{d}$ is the dust temperature of FIR galaxies. In this work, we fix $T_{d} = 34\rm{K}$ and $\beta = 2$ following \cite{Hall10}.

We make use of the MCMC method to analyze the parameters of CIB model by MontePython \citep{Audren13}. The likelihood is calculated by $\mathcal{L}=\exp(-\chi^2_{\rm tot}/2)$. We calculate the $\chi^2$ value using both the CIB auto and $\rm CIB\times CMB$ lensing angular power spectra at 250, 350, and 500 $\rm{\mu m}$, so that $\chi^2_{\rm tot}=\chi^2_{\rm auto}+\chi^2_{\rm cross}$. Since the relative statistical weight mainly can be reflected by the SNR of the data, the ratio of that  is about 2:1 for the auto and cross power spectra in our fitting process. The $\chi^2$ is defined as,
\begin{equation} \label{eq:chi2}
\chi^2=\sum_{i=1}^{N} \left(\frac{C_{\ell_i}^{\rm obs}-C_{\ell_i}^{\rm th}}{\sigma_{\ell_i}^{\rm obs}}\right)^2,
\end{equation}
where $N$ is the number of data points, $C_{\ell_i}^{\rm obs}$ and $\sigma_{\ell_i}^{\rm obs}$ are the observed power spectrum and error from observation at $i$-th bin, respectively, and $C_{\ell_i}^{\rm th}$ are the theoretical power spectrum. Note that the $C_{\ell}$ may not be Gaussian distributed when the SNR is low, and the chi-square statistics can be unavailable in such case. However, since the SNRs are $\sim$15 and $\sim$7 for the auto and cross power spectra, respectively, in our analysis, which are large enough, we still use the chi-square method shown in Eq.~(\ref{eq:chi2}) as a good approximation.

We constrain the parameter $A$, $z_c$, $\sigma_z$, DGL amplitude $A_{\rm DGL}$ and shot noise for each band, so the parameter space have 15 free parameters. We set the flat priors of the CIB model parameters: $A$, $A_{\rm DGL}$ and shot noise $\in (0,+\infty)$, and $z_c$ and $\sigma_z \in (0,6)$. For each case, we run 50 chains, and each chain contains 500,000 steps. After thinning the chains, we obtain about 15,000 chain points to illustrate the probability distribution of the parameters.

\begin{table}[ht!]
\begin{center}
\caption{\label{tab:result} The best-fits and errors of the parameters in the CIB model from the MCMC fitting.}
\small
\begin{tabular}{c c c c}
\hline\hline
\rule[-2mm]{0mm}{6mm}
Parameter & $250\rm{\mu m}$ & $350\rm{\mu m}$ & $500\rm{\mu m}$\\
\hline
\rule[-2mm]{0mm}{6mm}
$A\times10^{41}$            &$3.12^{+0.19}_{-0.47}$&$4.89^{+0.32}_{-0.41}$&$8.30^{+1.57}_{-2.02}$\\
\rule[-2mm]{0mm}{6mm}
$z_c$                                  &$1.52^{+0.47}_{-0.41}$&$1.48^{+0.41}_{-0.20}$&$1.14^{+0.28}_{-0.25}$\\
\rule[-2mm]{0mm}{6mm}
$\sigma_z$                         &$0.74^{+0.38}_{-0.43}$&$0.42^{+0.09}_{-0.17}$&$0.21^{+0.03}_{-0.05}$\\
\rule[-2mm]{0mm}{6mm}
$A_{\rm DGL}\times10^{-12}$       &$26.95^{+1.79}_{-1.98}$&$6.27^{+0.56}_{-0.52}$&$0.91^{+0.09}_{-0.10}$\\
\rule[-2mm]{0mm}{6mm}
$\rm shot\ noise \times10^{-3}$    &$5.33^{+0.14}_{-0.15}$&$3.70^{+0.09}_{-0.09}$&$1.50^{+0.06}_{-0.06}$\\
\hline
\end{tabular}
\end{center}
\end{table}

\subsection{fitting result}
\label{fit_result}

\begin{figure}
\centering
\includegraphics[width=1.0\columnwidth]{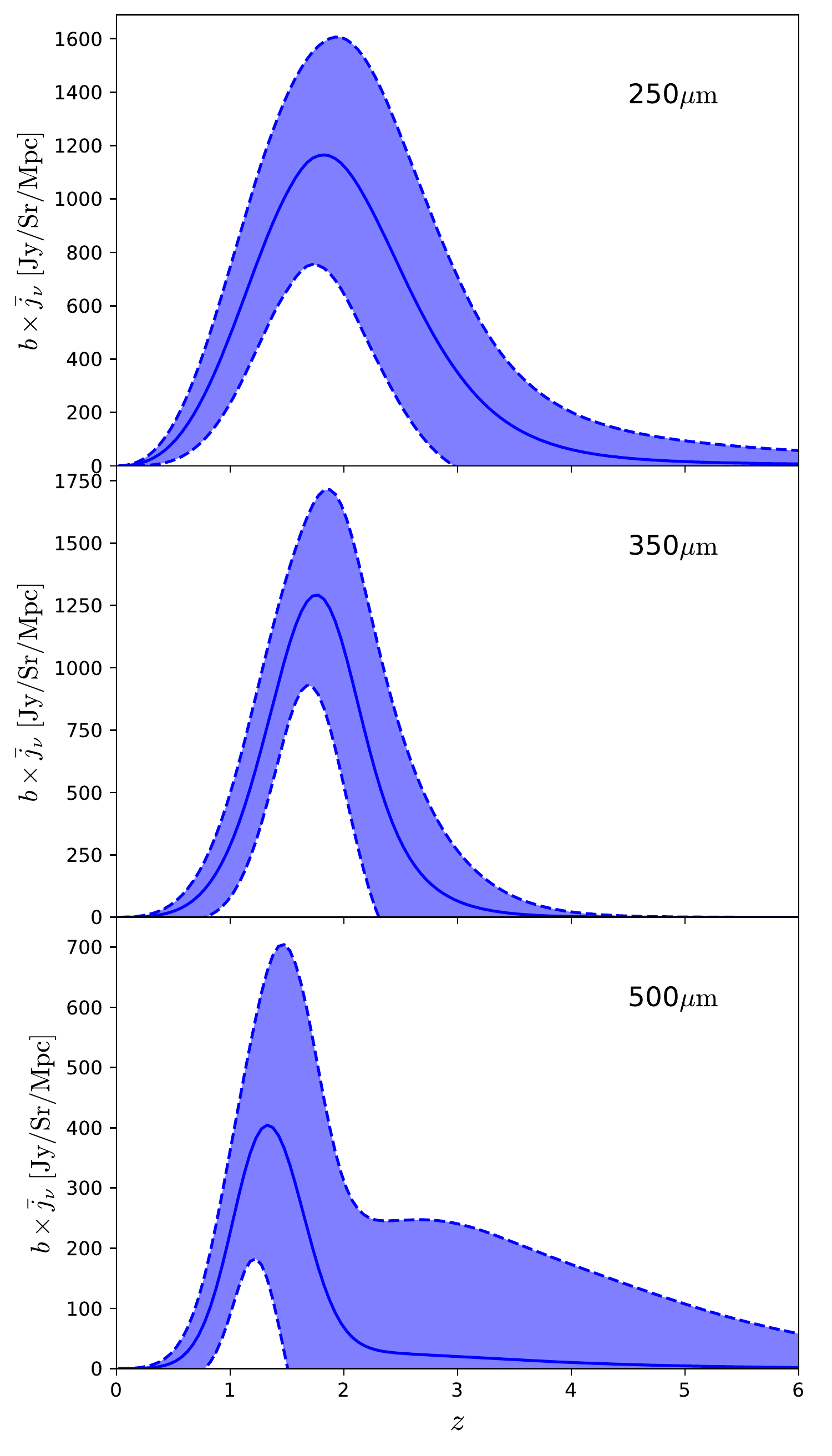}
\caption{The product of mean galaxy bias and CIB mean emissivity as a function of redshift derived from the MCMC chains at the three CIB bands. The solid curves and shaded regions show the mean values and standard deviations, respectively. Our result is basically consistent with \cite{Bethermin11} and \cite{Planck14XVIII} with much smaller uncertainties.}
\label{fig:j_nu} 
\end{figure}

In the right panel of Figure~\ref{fig:auto} and the Figure~\ref{fig:cross}, we show the best-fits of CIB auto and CIB$\times$CMB lensing power spectra in orange lines, and the reduced chi-square $\chi^2_{\rm red}=\chi^2_{\rm min}/N_{\rm dof}$, where $N_{\rm dof}$ is the degree of freedom, are 0.9, 0.8, and 0.7 for 250, 350, and 500 $\rm{\mu m}$, respectively. As can be seen, we can fit the data quite well, and the best-fit curves are consistent with most of the data points in 1-$\sigma$, especially for the CIB auto power spectra as shown in the right panel of Figure~{\ref{fig:auto}}. In this work, we estimate DGL using a simple model by asuming $C_{\ell}^{\rm{DGL}}= A_{\rm{DGL}}\ell^{-3}$ \citep{Mitchell15}, and fit the amplitude $A_{\rm{DGL}}$ based on the MCMC method. The large sky area used in this analysis is quite helpful for accurately deriving the DGL component. The best-fit values of $A_{\rm DGL}$ are found to be $2.70^{+0.18}_{-0.20} \times 10^{13}$, $6.27^{+0.56}_{-0.52} \times 10^{12}$, and $9.13^{+0.87}_{-1.04} \times 10^{11}$ at 250, 350, and 500 $\rm{\mu m}$, respectively as shown in Table~\ref{tab:result}. 

In Figure~\ref{fig:fit}, the contour maps of $z_c$ and $\sigma_z$ with 1-$\sigma$ and 2-$\sigma$ confidence levels (C.L.) are shown. The fitting results of $z_c$ and $\sigma_z$ are consistent with the results given by \cite{Planck14XVIII}, and we have much smaller uncertainties since the auto power spectra are included in out constraints. They did similar studies using the Planck HFI maps at 100, 143, 217, 353, 545, and 857 GHz in the analysis. The 545 GHz  (550 $\rm{\mu m}$) and 857 GHz (350 $\rm{\mu m}$) bands they used are similar to the 500~$\rm{\mu m}$ and 350 $\rm{\mu m}$ bands in Herschel survey, respectively. They find that $z_c=1$ and $\sigma_z=2.2$ have high probability at all frequencies, their uncertainty of $z_c$ and $\sigma_z$ are both greater than 1. Our constraints on $z_c$ and $\sigma_z$ are better than theirs by factors of $\sim$2 and $\sim$5,  respectively.

In Figure~\ref{fig:j_nu}, the CIB mean emissivities $\bar{j}_{\nu}(z)$ as a function of redshifts are shown for the three CIB bands, which are derived from the MCMC chains. We calculate $\bar{j}_{\nu}(z)$ for each chain point at $0<z<6$, and derive the mean values and standard deviations at the same redshift, which are shown in solid curves and shaded regions in Figure~\ref{fig:j_nu}, respectively. By comparing to previous studies, e.g. \cite{Bethermin11} and  \cite{Planck14XVIII}, we find that our results are basically in good agreements with theirs, and the uncertainties are significantly suppressed in our analysis.

\section{Summary}
\label{conclusions}
In this work, we have analyzed the auto- and cross-correlation of the cosmic far-infrared background of the Herschel SPIRE data and the cosmic microwave background lensing of Planck data in the Herschel-SPIRE HerMES Large Mode Survey field, which has 270 deg$^2$ and is the largest field in HerMES survey. The Herschel space telescope has three passbands that cover 250, 350, and 500 $\rm{\mu m}$. We adopt the Herschel Level 1 time stream data and the Planck second released data in our study.

First, we merged the maps using Madmap algorithm by HIPE, and set the pixel value of $6''$, $10''$ and $14''$ for 250, 350, and 500 $\rm{\mu m}$, respectively. Then we removed the pixels larger than $\rm{3\sigma (\sim 50mJy/beam)}$ and the areas that do not contain measured data. We considered different effects that can affect the measurements of the CIB power spectra, such as beam function, mode coupling, transfer function, and so on. Then we obtain the CIB power spectra at $100\le \ell \le 20,000$ for 250, 350, and 500 $\rm{\mu m}$. For the cross-correlation, we smoothed the Herschel maps to the resolution of CMB lensing map, and combined the masks of CIB maps and CMB lensing map to derive the cross-power spectra.

Next, we calculated and corrected the auto- and cross- power spectrum following \cite{Cooray12}. We obtained the instrumental noise by calculating the difference between of the auto- and cross-power spectrum of two scan measurements in the same band and the same field. We calculated the beam function by using a two-dimensional Gaussian beam, corrected the fictitious information from mask by a mode-coupling matrix, and simulated the impact of the map-marking process on the final power spectrum. We compare the results with the measurements from \cite{Planck14XXX} and \cite{Thacker13}, and find that the results are generally consistent.

Finally, we presented a linear bias model, with a normal distribution for the galaxy probability density. We performed a MCMC analysis on the CIB model parameters. We found that we can fit the data very well, with the reduced chi-squares less than 1 for all three CIB bands, especially for the CIB auto power spectra. The best fit value of $z_c$ is $1.52^{+0.47}_{-0.41}$, $1.48^{+0.41}_{-0.20}$ and $1.14^{+0.28}_{-0.25}$, and the best fit value of $\sigma_z$ is $0.74^{+0.38}_{-0.43}$, $0.42^{+0.09}_{-0.17}$ and $0.21^{+0.03}_{-0.05}$ for 250, 350, and 500 $\rm{\mu m}$, respectively. We also derived the corresponding CIB mean emissivity as a function of redshift from the MCMC chains for the three CIB bands. Our results are basically consistent with previous studies.

According to current results \citep{Amblard10,Cooray12,Thacker13,Viero13}, the far-infrared  background signal is dominated by the dusty star-forming galaxies with a redshift distribution peaked between z $\sim$ 1 and z $\sim$ 2. Our results are in good agreements with their predictions. According to the results of the auto- and cross-correlating, we are able to provide a check for the linear bias model. We find that such a model not only can fit the auto-correlations, but also can explain the cross-correlation signal. In the future work, we will use HOD models to explain the measured angular power spectra and extract more physical informations of the CIB.

\acknowledgments
YC and YG acknowledges the support of NSFC-11822305, NSFC-11773031, NSFC-11633004, MOST-2018YFE0120800, the Chinese Academy of Sciences (CAS) Strategic Priority Research Program XDA15020200, the NSFC-ISF joint research program No. 11761141012, and CAS Interdisciplinary Innovation Team.

\appendix
In Table~\ref{tab:auto} and Table~\ref{tab:cross}, the values of the CIB auto-power spectra and cross-power spectra of CIB and CMB Lensing at 250, 350, and 500 $\rm \mu m$ are shown, respectively.

\begin{table*}[ht!]
\begin{center}
\caption{\label{tab:auto} The CIB auto clustering power spectra at 250, 350, and 500 $\rm \mu m$.}
\small
\begin{tabular}{c c c c}
\hline\hline
\rule[-2mm]{0mm}{6mm}
 & $250\rm{\mu m}$ & $350\rm{\mu m}$ & $500\rm{\mu m}$\\
 \rule[-2mm]{0mm}{6mm}
$\ell$ & $\ell^2 C_{\ell}/2\pi \ [Jy^2/Sr^2]$ &$\ell^2 C_{\ell}/2\pi \ [Jy^2/Sr^2]$& $\ell^2 C_{\ell}/2\pi \ [Jy^2/Sr^2]$ \\
\hline
$1.15\times 10^2$&$(2.37+2.38)\times 10^{9}$&$(9.22+4.60)\times 10^8$&$(3.02+1.38)\times 10^8$\\
$1.50\times 10^2$&$(8.72+10.27)\times 10^{8}$&$(9.33+3.68)\times 10^8$&$(1.69+0.62)\times 10^8$\\
$1.96\times 10^2$&$(2.19+0.85)\times 10^{9}$&$(5.99+2.07)\times 10^8$&$(2.40+0.59)\times 10^8$\\
$2.55\times 10^2$&$(4.77+0.96)\times 10^{9}$&$(1.81+0.30)\times 10^9$&$(4.13+0.64)\times 10^8$\\
$3.32\times 10^2$&$(3.41+0.62)\times 10^{9}$&$(1.41+0.21)\times 10^9$&$(3.42+0.51)\times 10^8$\\
$4.33\times 10^2$&$(6.02+0.83)\times 10^{9}$&$(2.30+0.28)\times 10^9$&$(5.51+0.62)\times 10^8$\\
$5.64\times 10^2$&$(5.45+0.59)\times 10^{9}$&$(2.25+0.22)\times 10^9$&$(4.78+0.52)\times 10^8$\\
$7.36\times 10^2$&$(5.70+0.56)\times 10^9$&$(2.04+0.19)\times 10^9$&$(4.72+0.50)\times 10^8$\\
$9.59\times 10^2$&$(5.28+0.50)\times 10^9$&$(2.28+0.21)\times 10^9$&$(6.41+0.70)\times 10^8$\\
$1.25\times 10^3$&$(5.67+0.50)\times 10^9$&$(2.60+0.23)\times 10^9$&$(8.95+0.92)\times 10^8$\\
$1.63\times 10^3$&$(6.13+0.53)\times 10^9$&$(3.10+0.27)\times 10^9$&$(9.43+1.08)\times 10^8$\\
$2.12\times 10^3$&$(7.06+0.62)\times 10^9$&$(3.61+0.38)\times 10^9$&$(1.01+0.14)\times 10^9$\\
$2.77\times 10^3$&$(7.73+0.73)\times 10^9$&$(4.46+0.44)\times 10^9$&$(1.43+0.20)\times 10^9$\\
$3.61\times 10^3$&$(9.70+0.99)\times 10^{9}$&$(4.97+0.57)\times 10^9$&$(1.74+0.29)\times 10^9$\\
$4.70\times 10^3$&$(9.96+1.26)\times 10^{9}$&$(5.66+0.79)\times 10^9$&$(2.11+0.44)\times 10^9$\\
$6.12\times 10^3$&$(1.13+0.18)\times 10^{10}$&$(6.69+1.14)\times 10^9$&$(2.88+0.69)\times 10^9$\\
$7.98\times 10^3$&$(1.33+0.27)\times 10^{10}$&$(8.12+1.72)\times 10^9$&$(3.21+1.07)\times 10^9$\\
$1.04\times 10^4$&$(1.62+0.41)\times 10^{10}$&$(9.67+2.66)\times 10^{9}$&$(3.78+1.70)\times 10^9$\\
$1.36\times 10^4$&$(1.96+0.64)\times 10^{10}$&$(1.17+0.42)\times 10^{10}$&$(3.68+2.72)\times 10^9$\\
$1.77\times 10^4$&$(2.65+1.04)\times 10^{10}$&$(1.37+0.67)\times 10^{10}$&$(4.53+3.52)\times 10^9$\\
\hline
\end{tabular}
\normalsize
\end{center}
\end{table*}

\begin{table}[ht!]
\begin{center}
\caption{\label{tab:cross} The cross power spectra of CIB and CMB Lensing  at 250, 350, and 500 $\rm \mu m$.}
\small
\begin{tabular}{c c c c}
\hline\hline
\rule[-2mm]{0mm}{6mm}
 & $250\rm{\mu m}$ & $350\rm{\mu m}$ & $500\rm{\mu m}$\\
 \rule[-2mm]{0mm}{6mm}
$\ell$ & $\ell^3 C_{\ell} \ [Jy/Sr]$ &$\ell^3 C_{\ell} \ [Jy/Sr]$& $\ell^3 C_{\ell} \ [Jy/Sr]$ \\
\hline
$181$&$73.26\pm6.98$&$48.03\pm4.52$&$28.12\pm2.47$\\
$343$&$57.85\pm8.43$&$46.38\pm4.82$&$23.04\pm2.55$\\
$506$&$26.51\pm7.99$&$25.21\pm4.45$&$18.40\pm2.34$\\
$668$&$31.48\pm8.02$&$26.62\pm4.53$&$9.43\pm2.40$\\
$830$&$19.67\pm8.06$&$21.44\pm4.49$&$11.33\pm2.39$\\
$993$&$31.33\pm8.33$&$30.88\pm4.66$&$11.26\pm2.50$\\
$1155$&$35.02\pm8.58$&$19.31\pm4.86$&$6.93\pm2.63$\\
$1317$&$33.59\pm9.41$&$37.36\pm5.35$&$7.92\pm2.93$\\
$1480$&$13.05\pm10.00$&$28.47\pm5.75$&$9.15\pm3.17$\\
$1642$&$22.12\pm10.72$&$13.13\pm6.21$&$19.27\pm3.45$\\
$1804$&$2.92\pm10.94$&$16.03\pm6.40$&$2.29\pm3.58$\\
$1967$&$32.28\pm11.10$&$13.88\pm6.56$&$13.08\pm3.69$\\
\hline
\end{tabular}
\normalsize
\end{center}
\end{table}

\end{document}